 \message
 {JNL.TEX version 0.92 as of 6/9/87.  Report bugs and problems to Doug Eardley.}
 
 \catcode`@=11
 \expandafter\ifx\csname inp@t\endcsname\relax\let\inp@t=\input
 \def\input#1 {\expandafter\ifx\csname #1IsLoaded\endcsname\relax
 \inp@t#1%
 \expandafter\def\csname #1IsLoaded\endcsname{(#1 was previously loaded)}
 \else\message{\csname #1IsLoaded\endcsname}\fi}\fi
 \catcode`@=12

 
 
 \font\twelverm=cmr10 scaled 1200    \font\twelvei=cmmi10 scaled 1200
 \font\twelvesy=cmsy10 scaled 1200   \font\twelveex=cmex10 scaled 1200
 \font\twelvebf=cmbx10 scaled 1200   \font\twelvesl=cmsl10 scaled 1200
 \font\twelvett=cmtt10 scaled 1200   \font\twelveit=cmti10 scaled 1200
 \font\twelvesc=cmcsc10 scaled 1200  \font\twelvesf=cmssdc10 scaled 1200
 
 
 \def\twelvepoint{\normalbaselineskip=12.4pt plus 0.1pt minus 0.1pt
   \abovedisplayskip 12.4pt plus 3pt minus 9pt
   \abovedisplayshortskip 0pt plus 3pt
   \belowdisplayshortskip 7.2pt plus 3pt minus 4pt
   \smallskipamount=3.6pt plus1.2pt minus1.2pt
   \medskipamount=7.2pt plus2.4pt minus2.4pt
   \bigskipamount=14.4pt plus4.8pt minus4.8pt
   \def\rm{\fam0\twelverm}          \def\it{\fam\itfam\twelveit}%
   \def\sl{\fam\slfam\twelvesl}     \def\bf{\fam\bffam\twelvebf}%
   \def\mit{\fam 1}                 \def\cal{\fam 2}%
   \def\sc{\twelvesc}               \def\tt{\twelvett}
   \def\sf{\twelvesf}
   \textfont0=\twelverm   \scriptfont0=\tenrm   \scriptscriptfont0=\sevenrm
   \textfont1=\twelvei    \scriptfont1=\teni    \scriptscriptfont1=\seveni
   \textfont2=\twelvesy   \scriptfont2=\tensy   \scriptscriptfont2=\sevensy
   \textfont3=\twelveex   \scriptfont3=\twelveex  \scriptscriptfont3=\twelveex
   \textfont\itfam=\twelveit
   \textfont\slfam=\twelvesl
   \textfont\bffam=\twelvebf \scriptfont\bffam=\tenbf
   \scriptscriptfont\bffam=\sevenbf
   \normalbaselines\rm}
 

 
 \def\beginlinemode{\endmode
   \begingroup\parskip=0pt \obeylines\def\\{\par}\def\endmode{\par\endgroup}}
 \def\beginparmode{\endmode
   \begingroup \def\endmode{\par\endgroup}}
 \let\endmode=\par
 {\obeylines\gdef\
 {}}
 \def\singlespace{\baselineskip=\normalbaselineskip}
 
 \def\oneandahalfspace{\baselineskip=\normalbaselineskip
   \multiply\baselineskip by 3 \divide\baselineskip by 2}
 \def\doublespace{\baselineskip=\normalbaselineskip \multiply\baselineskip by 2}

 \newcount\firstpageno
 \firstpageno=2
 \footline={\ifnum\pageno<\firstpageno{\hfil}\else{\hfil\twelverm\folio
  \hfil}\fi}
 \def\toppageno{\global\footline={\hfil}\global\headline
   ={\ifnum\pageno<\firstpageno{\hfil}\else{\hfil\twelverm\folio\hfil}
  \fi}}
 \let\rawfootnote=\footnote              
 \def\footnote#1#2{{\rm\singlespace\parindent=0pt\parskip=0pt
   \rawfootnote{#1}{#2\hfill\vrule height 0pt depth 6pt width 0pt}}}
 \def\raggedcenter{\leftskip=4em plus 12em \rightskip=\leftskip
   \parindent=0pt \parfillskip=0pt \spaceskip=.3333em \xspaceskip=.5em
   \pretolerance=9999 \tolerance=9999
   \hyphenpenalty=9999 \exhyphenpenalty=9999 }
 \def\dateline{\rightline{\ifcase\month\or
   January\or February\or March\or April\or May\or June\or
   July\or August\or September\or October\or November\or December\fi
   \space\number\year}}
 \def\received{\vskip 3pt plus 0.2fill
  \centerline{\sl (Received\space\ifcase\month\or
   January\or February\or March\or April\or May\or June\or
   July\or August\or September\or October\or November\or December\fi
   \qquad, \number\year)}}
 
 
 \hsize=6.5truein
 \hoffset=0pt
 \vsize=8.9truein
 \voffset=0pt
 \parskip=\medskipamount
 \def\\{\cr}
 \twelvepoint            
 \doublespace            
 \overfullrule=0pt       
 
 
 
 
 \def\title                      
   {\null\vskip 3pt plus 0.2fill
    \beginlinemode \doublespace \raggedcenter \bf}
 
 \def\author                     
   {\vskip 3pt plus 0.2fill \beginlinemode
    \singlespace \raggedcenter\sl}    
 
 \def\affil                      
   {\vskip 3pt plus 0.1fill \beginlinemode
    \oneandahalfspace \raggedcenter \sl}
 
 \def\abstract                   
   {\vskip 3pt plus 0.3fill \beginparmode
    \oneandahalfspace ABSTRACT: }
 
 \def\endtitlepage               
   {\endpage                     
    \body}
 \let\endtopmatter=\endtitlepage
 
 \def\body                       
   {\beginparmode}               
 
 \def\head#1{                    
   \goodbreak\vskip 0.5truein    
   {\immediate\write16{#1}
    \raggedcenter \uppercase{#1}\par}
    \nobreak\vskip 0.25truein\nobreak}

 \def\beginitems{
 \par\medskip\bgroup\def\i##1 {\item{##1}}\def\ii##1 {\itemitem{##1}}
 \leftskip=36pt\parskip=0pt}
 \def\enditems{\par\egroup}
 
 \def\beneathrel#1\under#2{\mathrel{\mathop{#2}\limits_{#1}}}

 \def\refto#1{[#1]}
 
 \def\references                 
   {\head{References}            
    \beginparmode
    \frenchspacing \parindent=0pt \leftskip=1truecm
    \parskip=8pt plus 3pt \everypar{\hangindent=\parindent}}

 \gdef\refis#1{\item{#1.\ }}                     

 \gdef\journal#1, #2, #3, 1#4#5#6{           
 {\sl #1~}{\bf #2}, #3 (1#4#5#6)}            


 \def\endreferences{\body}
 
 \def\figurecaptions             
   {\endpage
    \beginparmode
    \head{Figure Captions}
 }
 
 \def\endfigurecaptions{\body}
 
 \def\endpage                    
   {\vfill\eject}
 
 \def\endpaper                   
   {\endmode\vfill\supereject}


 \def\heading                            
   {\vskip 0.5truein plus 0.1truein      
    \beginparmode \def\\{\par} \parskip=0pt \singlespace \raggedcenter}

 \def\subheading                         
   {\vskip 0.25truein plus 0.1truein     
    \beginlinemode \singlespace \parskip=0pt \def\\{\par}\raggedcenter}

 \def\tag#1$${\eqno(#1)$$}
 
 \def\align#1$${\eqalign{#1}$$}

 \def\aligntag#1$${\gdef\tag##1\\{&(##1)\cr}\eqalignno{#1\\}$$
   \gdef\tag##1$${\eqno(##1)$$}}
 
 \def\endaligntag{}

 \def\overset #1\to#2{{\mathop{#2}\limits^{#1}}}
 \def\underset#1\to#2{{\let\next=#1\mathpalette\undersetpalette#2}}
 \def\undersetpalette#1#2{\vtop{\baselineskip0pt
 \ialign{$\mathsurround=0pt #1\hfil##\hfil$\crcr#2\crcr\next\crcr}}}

 
 \def\ref#1{Ref.~#1}                     
 \def\Ref#1{Ref.~#1}                     
 \def\[#1]{[\cite{#1}]}
 \def\cite#1{{#1}}
 \def\(#1){(\call{#1})}
 \def\call#1{{#1}}
 \def\taghead#1{}
 \def\frac#1#2{{#1 \over #2}}
 \def\half{{\frac 12}}

 \def\12{{1\over2}}

 \def\sla{\raise.15ex\hbox{$/$}\kern-.57em}
 \def\leaderfill{\leaders\hbox to 1em{\hss.\hss}\hfill}
 \def\twiddle{\lower.9ex\rlap{$\kern-.1em\scriptstyle\sim$}}
 \def\bigtwiddle{\lower1.ex\rlap{$\sim$}}
 \def\gtwid{\mathrel{\raise.3ex\hbox{$>$\kern-.75em\lower1ex\hbox{$\sim$}}}}
 \def\ltwid{\mathrel{\raise.3ex\hbox{$<$\kern-.75em\lower1ex\hbox{$\sim$}}}}
 \def\square{\kern1pt\vbox{\hrule height 1.2pt\hbox{\vrule width 1.2pt\hskip 3pt
    \vbox{\vskip 6pt}\hskip 3pt\vrule width 0.6pt}\hrule height 0.6pt}\kern1pt}
 \def\tdot#1{\mathord{\mathop{#1}\limits^{\kern2pt\ldots}}}
 
 \def\pmb#1{\setbox0=\hbox{#1}%
   \kern-.025em\copy0\kern-\wd0
   \kern  .05em\copy0\kern-\wd0
   \kern-.025em\raise.0433em\box0 }

\catcode`@=11
\newcount\tagnumber\tagnumber=0

\immediate\newwrite\eqnfile
\newif\if@qnfile\@qnfilefalse
\def\write@qn#1{}
\def\writenew@qn#1{}
\def\w@rnwrite#1{\write@qn{#1}\message{#1}}
\def\@rrwrite#1{\write@qn{#1}\errmessage{#1}}

\def\taghead#1{\gdef\t@ghead{#1}\global\tagnumber=0}
\def\t@ghead{}

\expandafter\def\csname @qnnum-3\endcsname
  {{\t@ghead\advance\tagnumber by -3\relax\number\tagnumber}}
\expandafter\def\csname @qnnum-2\endcsname
  {{\t@ghead\advance\tagnumber by -2\relax\number\tagnumber}}
\expandafter\def\csname @qnnum-1\endcsname
  {{\t@ghead\advance\tagnumber by -1\relax\number\tagnumber}}
\expandafter\def\csname @qnnum0\endcsname
  {\t@ghead\number\tagnumber}
\expandafter\def\csname @qnnum+1\endcsname
  {{\t@ghead\advance\tagnumber by 1\relax\number\tagnumber}}
\expandafter\def\csname @qnnum+2\endcsname
  {{\t@ghead\advance\tagnumber by 2\relax\number\tagnumber}}
\expandafter\def\csname @qnnum+3\endcsname
  {{\t@ghead\advance\tagnumber by 3\relax\number\tagnumber}}

\def\equationfile{%
  \@qnfiletrue\immediate\openout\eqnfile=\jobname.eqn%
  \def\write@qn##1{\if@qnfile\immediate\write\eqnfile{##1}\fi}
  \def\writenew@qn##1{\if@qnfile\immediate\write\eqnfile
    {\noexpand\tag{##1} = (\t@ghead\number\tagnumber)}\fi}
}

\def\callall#1{\xdef#1##1{#1{\noexpand\call{##1}}}}
\def\call#1{\each@rg\callr@nge{#1}}

\def\each@rg#1#2{{\let\thecsname=#1\expandafter\first@rg#2,\end,}}
\def\first@rg#1,{\thecsname{#1}\apply@rg}
\def\apply@rg#1,{\ifx\end#1\let\next=\relax%
\else,\thecsname{#1}\let\next=\apply@rg\fi\next}

\def\callr@nge#1{\calldor@nge#1-\end-}
\def\callr@ngeat#1\end-{#1}
\def\calldor@nge#1-#2-{\ifx\end#2\@qneatspace#1 %
  \else\calll@@p{#1}{#2}\callr@ngeat\fi}
\def\calll@@p#1#2{\ifnum#1>#2{\@rrwrite{Equation range #1-#2\space is bad.}
\errhelp{If you call a series of equations by the notation M-N, then M and
N must be integers, and N must be greater than or equal to M.}}\else%
 {\count0=#1\count1=#2\advance\count1 by1\relax\expandafter\@qncall\the\count0,%
  \loop\advance\count0 by1\relax%
    \ifnum\count0<\count1,\expandafter\@qncall\the\count0,%
  \repeat}\fi}

\def\@qneatspace#1#2 {\@qncall#1#2,}
\def\@qncall#1,{\ifunc@lled{#1}{\def\next{#1}\ifx\next\empty\else
  \w@rnwrite{Equation number \noexpand\(>>#1<<) has not been defined yet.}
  >>#1<<\fi}\else\csname @qnnum#1\endcsname\fi}

\let\eqnono=\eqno
\def\eqno(#1){\tag#1}
\def\tag#1$${\eqnono(\displayt@g#1 )$$}

\def\aligntag#1\endaligntag
  $${\gdef\tag##1\\{&(##1 )\cr}\eqalignno{#1\\}$$
  \gdef\tag##1$${\eqnono(\displayt@g##1 )$$}}

\def\eqalignno#1{\displ@y \tabskip\centering
  \halign to\displaywidth{\hfil$\displaystyle{##}$\tabskip\z@skip
    &$\displaystyle{{}##}$\hfil\tabskip\centering
    &\llap{$\displayt@gpar##$}\tabskip\z@skip\crcr
    #1\crcr}}

\def\displayt@gpar(#1){(\displayt@g#1 )}

\def\displayt@g#1 {\rm\ifunc@lled{#1}\global\advance\tagnumber by1
        {\def\next{#1}\ifx\next\empty\else\expandafter
        \xdef\csname @qnnum#1\endcsname{\t@ghead\number\tagnumber}\fi}%
  \writenew@qn{#1}\t@ghead\number\tagnumber\else
        {\edef\next{\t@ghead\number\tagnumber}%
        \expandafter\ifx\csname @qnnum#1\endcsname\next\else
        \w@rnwrite{Equation \noexpand\tag{#1} is a duplicate number.}\fi}%
  \csname @qnnum#1\endcsname\fi}

\def\ifunc@lled#1{\expandafter\ifx\csname @qnnum#1\endcsname\relax}

\let\@qnend=\end\gdef\end{\if@qnfile
\immediate\write16{Equation numbers written on []\jobname.EQN.}\fi\@qnend}

\catcode`@=12

 \catcode`@=11
 \newcount\r@fcount \r@fcount=0
 \newcount\r@fcurr
 \immediate\newwrite\reffile
 \newif\ifr@ffile\r@ffilefalse
 \def\w@rnwrite#1{\ifr@ffile\immediate\write\reffile{#1}\fi\message{#1}}
 
 \def\writer@f#1>>{}
 \def\referencefile{
   \r@ffiletrue\immediate\openout\reffile=\jobname.ref%
   \def\writer@f##1>>{\ifr@ffile\immediate\write\reffile%
     {\noexpand\refis{##1} = \csname r@fnum##1\endcsname = %
      \expandafter\expandafter\expandafter\strip@t\expandafter%
      \meaning\csname r@ftext\csname r@fnum##1\endcsname\endcsname}\fi}%
   \def\strip@t##1>>{}}

 \def\citeall#1{\xdef#1##1{#1{\noexpand\cite{##1}}}}
 \def\cite#1{\each@rg\citer@nge{#1}}     
 
 \def\each@rg#1#2{{\let\thecsname=#1\expandafter\first@rg#2,\end,}}
 \def\first@rg#1,{\thecsname{#1}\apply@rg}       
 \def\apply@rg#1,{\ifx\end#1\let\next=\relax
 \else,\thecsname{#1}\let\next=\apply@rg\fi\next}
 
 \def\citer@nge#1{\citedor@nge#1-\end-}  
 \def\citer@ngeat#1\end-{#1}
 \def\citedor@nge#1-#2-{\ifx\end#2\r@featspace#1 
   \else\citel@@p{#1}{#2}\citer@ngeat\fi}        
 \def\citel@@p#1#2{\ifnum#1>#2{\errmessage{Reference range #1-#2\space is bad}.%
     \errhelp{If you cite a series of references by the notation M-N, then M and
     N must be integers, and N must be greater than or equal to M.}}\else
  {\count0=#1\count1=#2\advance\count1 by1\relax\expandafter\r@fcite\the\count0,%
   \loop\advance\count0 by1\relax
     \ifnum\count0<\count1,\expandafter\r@fcite\the\count0,%
   \repeat}\fi}
 
 \def\r@featspace#1#2 {\r@fcite#1#2,}    
 \def\r@fcite#1,{\ifuncit@d{#1}
     \newr@f{#1}%
     \expandafter\gdef\csname r@ftext\number\r@fcount\endcsname%
                      {\message{Reference #1 to be supplied.}%
                       \writer@f#1>>#1 to be supplied.\par}%
  \fi%
  \csname r@fnum#1\endcsname}
 \def\ifuncit@d#1{\expandafter\ifx\csname r@fnum#1\endcsname\relax}%
 \def\newr@f#1{\global\advance\r@fcount by1%
     \expandafter\xdef\csname r@fnum#1\endcsname{\number\r@fcount}}
 
 \let\r@fis=\refis                       
 \def\refis#1#2#3\par{\ifuncit@d{#1}
 blank
    \newr@f{#1}%
    \w@rnwrite{Reference #1=\number\r@fcount\space is not cited up to now.}\fi%
   \expandafter\gdef\csname r@ftext\csname r@fnum#1\endcsname\endcsname%
   {\writer@f#1>>#2#3\par}}
 
 \def\ignoreuncited{
    \def\refis##1##2##3\par{\ifuncit@d{##1}%
      \else\expandafter\gdef\csname r@ftext\csname r@fnum##1\endcsname\endcsname%
      {\writer@f##1>>##2##3\par}\fi}}
 
 \def\r@ferr{\endreferences\errmessage{I was expecting to see
 \noexpand\endreferences before now;  I have inserted it here.}}
 \let\r@ferences=\references
 \def\references{\r@ferences\def\endmode{\r@ferr\par\endgroup}}
 
 \let\endr@ferences=\endreferences
 \def\endreferences{\r@fcurr=0
   {\loop\ifnum\r@fcurr<\r@fcount
     \advance\r@fcurr by 1\relax\expandafter\r@fis\expandafter{\number\r@fcurr}%
     \csname r@ftext\number\r@fcurr\endcsname%
   \repeat}\gdef\r@ferr{}\endr@ferences}
 
 
 \let\r@fend=\endpaper\gdef\endpaper{\ifr@ffile
 \immediate\write16{Cross References written on []\jobname.REF.}\fi\r@fend}
 
 \catcode`@=12

 \citeall\refto          
 \citeall\ref            %
 \citeall\Ref            %
 

\newcount\notenumber
\def\clearnotenumber{\notenumber=0}
\def\note{\advance\notenumber by1 \footnote{$^{\the\notenumber}$}}
\clearnotenumber
\def\ch{\mathaccent 94}
\def\til{\mathaccent "7E }

\def \r0{\ch{\rho}^{(0)}}
\def\rr{ {\bf r} }
\def\RR{ {\bf R} }
\def\TT{ {\bf T} }
\def\HH{ {\bf H} }
\def\FF{ {\bf F} }
\def\qq{ {\bf q} }
\def\kk{ {\bf k} }
\def\FF{ {\bf F} }
\def\EE{ {\bf E} }
\def\VV{ {\bf V} }
\def\UU{ {\bf U} }
\def\vv{ {\bf v} }
\def\uu{ {\bf u} }
\def\ttt{ {\bf t} }

\def\q0{q_{_0}}

\def\lb{\ell_{_B}}
\def\l0{\ell_{_0}}
\def\lo{\ell_{_{OSF}}}

\title THEORY OF POLYELECTROLYTE SOLUTIONS

\author Jean-Louis BARRAT

\affil D\'epartement de Physique des Mat\'eriaux (URA CNRS 172),
Universit\'e Claude Bernard-Lyon I, 69622 Villeurbanne Cedex, France

\author Jean-Fran\c{c}ois JOANNY

\affil Institut Charles Sadron (UPR  CNRS 022), 6 rue Boussingault,
 67083 Strasbourg Cedex, France.

\vskip 1in
 to appear in {\it Advances in Chemical Physics} , 1995

\endpage
\endtopmatter

\singlespace

\centerline{\bf TABLE OF CONTENTS}

\parindent=0pt

1. Introduction 

2. Charged chains at infinite dilution - asymptotic properties

\parindent=30pt

 {\it 2.1 Definition of the model and Flory-like calculation}

 {\it 2.2 Variational approaches}

 {\it 2.3  Renormalization group calculations}

 {\it 2.4  Screening of electrostatic interactions}

 {\it 2.5  Annealed and quenched polyelectrolytes}

\parindent=0pt

3. Local aspects of screening

\parindent=30pt

{\it 3.1 Counterion condensation}

{\it 3.2  Poisson Boltzmann approach}

{\it 3.3  Attractive electrostatic interactions}

\parindent=0pt

4. Electrostatic rigidity

\parindent=30pt

{\it 4.1  The Odijk-Skolnick-Fixman theory}

{\it 4.2 Alternative calculations for flexible chains}

{\it 4.3  The case of poor solvents}

\parindent=0pt

5. Charged gels and brushes

\parindent=30pt

{\it 5.1 Grafted polyelectrolyte layers} 

{\it 5.2 Polyelectrolyte gels}

\parindent=0pt

6. Semidilute solutions

\parindent=30pt

{\it 6.1 Ordering transitions in polyelectrolyte solutions}

{\it 6.2 Correlation length and osmotic pressure of semi-dilute polyelectrolyte solutions}

{\it 6.3 Electrostatic rigidity in semidilute solutions}

{\it 6.4 Concentrated  solutions of flexible polyelectrolytes}

{\it 6.5 Mesophase formation in poor solvents}

\parindent=0pt

7. Dynamical properties
\parindent=30pt

{\it 7.1 Mobility and electrophoretic mobility of a single charged chain}

{\it 7.2 Viscosity of polyelectrolyte solutions}

\parindent 0pt

8. Conclusions

Appendix A: effective interaction between charged monomers

Appendix B: relaxation and electrophoretic effect

\parindent=20pt

\endpage

\doublespace
{\bf 1. Introduction}

\smallskip

Polyelectrolytes are polymers bearing ionizable groups,
which, in polar solvents, can dissociate into charged 
polymer chains (macro-ions) and small "counter-ions".
Well known examples of such systems are proteins,
nucleic acids or synthetic systems such as sulfonated
polystyrene or polyacrylicacid. Very often, the polar solvent is water. 
Because of their fundamental importance
in biology and biochemistry, and because of their hydrosolubility,
ionizable polymers have been the object of a continued interest
since the early days of polymer science (see e.g. \refto{TA61,Mandel}
and references therein). It is however, a widely acknowledged fact
\refto{Schmitz} that they 
remain among the least understood systems in macromolecular science.
This situation is in sharp contrast with the 
case of neutral polymers.  In fact,
even  the properties that are now routinely used
to {\it characterize} neutral polymers (e.g. light scattering,
osmometry, viscosity..) are still very  poorly understood
in polyelectrolyte solutions.

Briefly speaking, the origin of this difference with
neutral polymers
can be traced back to the difficulty 
to apply renormalization group theories and scaling ideas to systems in which 
long range (coulomb)  forces are present. 
The success of the modern theories (and, although this was
not always realized as they were developped,  of many
older approaches) of neutral polymer solutions
is based on the fact that the range of the interactions 
between molecules is much smaller than the scale
determining the physical properties of the solution, 
which is  the size of the polymer chain or the correlation length.
If, as it is in general the case in polymer physics, the main
issue is the variation of solution properties with molecular
weight, the interactions only
affect  prefactors that can be adjusted 
to experimental results. The theories \refto{GE79,DJ85,DE86}
developped in this spirit have proven extremely
powerful for the interpretation of experimental data 
on neutral polymer solutions. Polyelectrolyte
solutions are more complex, both short range (excluded volume)
and long range (coulombic) interactions are simultaneously
present. The screening of the electrostatic interactions
introduces an intermediate length scale in the problem, that can
be comparable to the chain size or to the correlation length.
Moreover, the details of the local chain structure are
important and control the  phenomenon of counterion condensation \refto{MA68}
 (see
section 3). This condensation in turn modifies the long range part of the
interaction. The long range interactions actually also has
have a nontrivial influence on the {\it local structure} 
  \refto{OD77,SF77,OH78} (stiffness) of the polymer
chains (section 4).
 The implication of this complicated coupling
between small and large length scales is that the theoretical results
for polyelectrolytes can be expected to be more "model-dependent"
than for neutral polymers. In particular, the functional dependence
of several physical properties on molecular weight or concentration
depends in some cases on the local description 
that is used for the polymer chain. Such situations
never occur in neutral polymers.

Intrinsically related to this theoretical problem, is the more
practical difficulty in comparing experiments and theory
for polyelectrolyte solutions. In neutral polymers,
the comparison is made possible by adjusting 
interaction dependent prefactors and fitting experimental results
to the theoretically predicted dependence on, e.g. molecular
weight. In polyelectrolytes, "microscopic" parameters can 
have bigger influence,   and 
 a much more precise modelling might be necessary 
in order to interpret experimental results.  This often involves
the introduction of more adjustable parameters in the interpretation,
which sometimes becomes inconclusive.

In spite of these considerations,  the modern approach
to neutral polymers physics  has inspired a number of contributions 
to the polyelectrolyte problem in the last two decades. By "modern",
we mean here those theoretical approaches that attempt to use as much
as possible "coarse grained" models, with a minimal number of
microscopic parameters.
The goal of this review is to summarize these approaches, with the hope
that they could provide a useful conceptual framework for further
developpment. The review is organized as follows. Sections 2 to 4
are concerned with single chain properties. Section 2  deals 
with the academic problem of a charged chain in the absence of counterions. 
The phenomenon of screening, and the influence of the small
ions on the interaction between monomers, is discussed in section 3.
The influence of long range interaction on the local stiffness
(electrostatic rigidity) is described in section 4. 
In sections 5 and 6, the properties of interacting polyelectrolyte 
chains are considered. The simpler case of gels and brushes -in which
the structure is controlled by the preparation of the system- is described
first. Semidilute and concentrated solutions are the subject of section 6.
Finally, some results on dynamical properties (diffusion, electrophoretic mobility and viscosity)
are presented in section
7. 

Some important subjects have been left out of the scope of this review,
either because they have been reviewed recently elsewhere or 
because our understanding is too fragmentary. Among those are 
the dielectric properties of polyelectrolyte solutions
\refto{OdijkMandel}, their rheological 
properties, computer simulation results \refto{Kremerdunweg},
and all the problems related to polyelectrolyte adsorption \refto{adsorption}.
Finally, we systematically refer 
the reader to the original papers for 
 detailed comparison with the experimental results. The reason, mentioned 
above, is that such comparisons are, in general, not simple,
and involve the determination of several parameters.

{\bf 2.  Charged chains at infinite dilution - asymptotic properties}

\taghead{2.}

\smallskip 

{\it 2.1 Definition of the Model and Flory-like calculation}

\smallskip

In the theoretical limit of infinite dilution and zero salt concentration, 
a polyelectrolyte chain 
can be modelled as a connected sequence of charged and uncharged monomers
in  a "dielectric vacuum" that represents the solvent.  This model is not very realistic
to describe actual polyelectrolyte solutions, as will become clear
in our discussion of ion condensation and screening but it 
serves here to introduce some important concepts relative to charged polymer chains.

The energy for such an isolated polymer chain, made up of $N$ monomers,
is written as:
$$\eqalign{
H = H_0 +{\half }k_B T \sum_{i=1,N} \sum_{j\neq i}{\lb z_i z_j \over 
\vert \rr_i-\rr_j \vert }
}\eqno(H)
$$
where $H_0$ is the energy of the neutral  polymer chain that 
would be obtained by "switching off"  the electrostatic interaction,
and the double summation on the left hand side includes all monomer pairs. 
Here $z_i$ is the charge (in units of the
 electronic charge $e$) of monomer $i$, $T$ is the 
absolute temperature and $\lb= e^2/(4\pi\epsilon_0 \epsilon_r k_B T)$
is the Bjerrum length, that characterizes the strength of
electrostatic interactions in the solvent. For water at $T=300K$, $\lb\simeq 0.7nm$.

The "neutral polymer" energy $H_0$ 
contains terms describing the chemical bonding of the monomers as well
as short range excluded volume terms. In this section, we focus
on the very simple case where $H_0$ is the free energy of a gaussian,
or bead-spring, 
chain, i.e.
$$\eqalign{
H_0 = { 3 k_BT \over 2 b^2} \sum_{i=1,N-1} (\rr_i-\rr_{i+1})^2\ .
}\eqno(H0)
$$
The "monomer size" $b$ is such that the mean squared averaged
end-to-end distance for the neutral chain is $R_0^2=Nb^2$.
The physical situation described by this model
is that of a very weakly charged,  flexible chain.  
In that case, each "spring" actually represents 
a flexible  subchain of several neutral monomers.
Moreover, the absence of excluded volume interactions
implies theta conditions for the neutral backbone. 
Although these conditions are rather unlikely to be
achieved in experiments, the model contains the two
key ingredients that monitor the asymptotic behavior
of charged polymers in the absence of screening,
 namely long range interactions and monomer connectivity. 
These are the only relevant contributions as can be seen 
from the form of the Flory free energy for a chain of $N$ monomers,
carrying a total charge $fNe$ (i.e. $f$ is the fraction of charged
monomers, $(1-f)N$ monomers are neutral). The Flory energy \refto{GP76}
 for a chain
\note{The Flory free energy ignores numerical prefactors of order unity;
the prefactor of the electrostatic energy depends on the actual distribution 
of the charges in the sphere of radius $R$, it is equal to $3/5$ 
if the charges are uniformly distributed}
of size $R$  is the sum of an elastic energy $k_BT R^2/Nb^2 $ 
and of a Coulombic energy $k_BT (Nf)^2 \lb/R$, 
$$\eqalign{
E_{Flory} = k_B T \left( { R^2 \over Nb^2 }+ {(Nf)^2 \lb \over R}\right) \ ,
}\eqno(flory)
$$
which upon minimization with respect to $R$ gives an equilibrium size
$$\eqalign{
R \sim  Nf^{2/3} (\lb b^2)^{1/3}\ . 
}\eqno(R)
$$
This establishes the well known result, that the chain size is proportional 
to the number of monomers and to a $1/3$ power of the charge
fraction $f$.  This result was first obtained  by Katchalsky and 
coworkers \refto{KK48} using a distribution function approach, shortly before 
Flory published an equivalent calculation for chains 
with short range excluded volume. 

The electrostatic interactions tend to swell the chain, therefore the equilibrium radius
of a weakly charged polyelectrolyte chain must be larger than the gaussian radius $R_0$. 
This condition defines a minimum charge fraction $f_G \simeq N^{-3/4} (b/lb)^{1/2}$ 
above which the electrostatic interactions are relevant and the radius is given by (\call{R}). 
If the fraction of charges $f$ is smaller than $f_G$ the polyelectrolyte chain 
essentially behaves as a neutral chain and has a gaussian radius $R_0$. In the 
following we always implicitly assume that $f>f_G$.

In contrast to Flory's calculation, whose success is known \refto{DJ85}
to arise from an uncontrolled cancellation of errors, Katchalsky's 
approach is now believed to yield an essentially 
exact scaling result, as will 
become clear in the next sections.  A more refined
calculation of the electrostatic energy, accounting for the fact 
that the shape of the chain is rodlike rather than spherical, only
yields a  logarithmic correction to (\call{R})
$$\eqalign{
R \sim  Nf^{2/3} (\lb b^2 \ln(N))^{1/3}  \ .
}\eqno(R1)
$$
Because of this rodlike character, the polyelectrolyte chain 
is often described as being a "fully extended" object. This statement,
however, somewhat overinterprets the actual meaning of relation
(\call{R}). Short range molecular flexibility does not have to be frozen
in a locally stretched configuration to obtain such relations.

To understand this point, and to get a deeper insight into the meaning
of Katchalsky's calculation, it is useful to introduce the concept
of electrostatic blob \refto{GP76}. An electrostatic blob is a chain subunit
within which the electrostatic interactions can be considered 
as a weak perturbation.  The size $\xi_e$
and the number of monomers $g$ in such a subunit are thus related by
$$\eqalign{
{(fg)^2  \lb \over \xi_e}\simeq 1
}\eqno(ee)
$$
Another relation between $\xi_e$ and $g$ is obtained from the
 gaussian statistics of the chains on the scale $\xi_e$, 
$\xi_e^2\simeq g b^2$. This  gives
$$\eqalign{
\xi_e &\sim  b (f^2\lb/b)^{-1/3}\cr
 g  &\sim (f^2\lb/b)^{-2/3}
}\eqno(blob)
$$
The size of the chain (equation (\call{R})) is therefore $(N/g)\xi_e$,
so that the chain can be thought of as a linear (in the sense that 
its lateral dimensions are vanishingly small compared  to its length)
string of "electrostatic blobs" (figure 1).

The blob picture is particularly useful
to discuss the structure of a charged chain in a good or
poor solvent. In a good solvent, the only modification compared to the 
theta solvent case is the good solvent statistics 
characterized by a swelling exponent $\nu \neq 1/2$
must be used in the relation that relates the blob size to the 
number of monomers in a blob, $\xi_e \sim g^{\nu}$. The result is
still a linear string of blobs, $R\sim N f^y$, but the exponent
$y$ that describes the variation with the charge fraction 
is now $(2-2\nu)/(2-\nu)$. In a poor solvent, the problem is more subtle,
and has been analysed in detail by Khokhlov \refto{KH80}. In that case, the
chain would be completely collapsed into a globular state
in the absence of electrostatic interactions. 
The effect of these interactions
is to break up the globular structure into a string of 
"globular blobs". The criterion that determines the blob size
is now that the electrostatic energy within the blob
is sufficient to overcome the interfacial energy that must be
paid when monomers are exposed to the solvent, i.e.
$$\eqalign{
{(fg)^2  \lb \over \xi_e}\sim   \Gamma \xi_e^2 / (k_BT)
}\eqno(cblob)
$$
where $\Gamma$ is the interfacial tension between the collapsed
phase and the pure solvent. A second relation is obtained by expressing 
that the blob has the same structure as a small section of collapsed 
polymer,
$$\eqalign{
 g \sim n \xi_e^3 \ .
}\eqno(cblob1)
$$
Here $n$ is the density of the collapsed chain. 
 In the simple case where the second and third virial coefficients between monomers
are, respectively, $b^3 (T-\theta)/T = - b^3 \tau$ and $b^6$,
$n\sim \tau b^{-3}$ and $\Gamma=k_B T \tau^2 b^{-2}$. Equations 
(\call{blob}) and (\call{R}) are replaced by 
$$\eqalign{
\xi_e & \sim b  (f^2  \lb /b )^{-1/3} \cr
 g    & \sim \tau (f^2  \lb /b )^{-1} \cr
 R    & \sim (N/g) \xi_e \sim N \tau^{-1}  (f^2  \lb /b )^{2/3}\ .
}\eqno(cblob2)
$$

\smallskip

{\it 2.2 Variational approaches }

The Flory-type calculation discussed above is a typical mean-field
approach. Similar approaches which have been successfully
applied to the charged bead spring model of section 2.1. 
include the "chain under tension" model \refto{GP76,BB93} and the 
"gaussian variational method" \refto{DJ85,BB93,BP92,JP95}. Both methods
are mean-field approaches, formulated in a variational fashion. 
The chain is approximated by a simpler system described
by a trial Hamiltonian $H_t$, and the "best" trial Hamiltonian is obtained
by minimizing the variational free energy
$$\eqalign{
F_{var} = <H>_t-T S_t
}\eqno(Fvar)
$$
where $<H>_t$ is the average value of the energy for the true system,
taken with a statistical weight $\exp(-H_t/k_BT)$, and
$S_t$ is the entropy of the trial system.

In the 'chain under tension` model, the trial
Hamiltonian is that of a noninteracting gaussian chain
subject to an external tension $\FF$, 
$$\eqalign{
H_t\left(\FF\right)={3k_BT \over 2b^2}\sum_{i=1}^{N-1} (\rr_{i+1}-\rr_i)^2\ +\FF.(\rr_N-\rr_1)\ .
}\eqno(CT)
$$
This model is simple enough to allow explicit calculations 
of many properties of the chain. In particular, it can be shown
that the minimization of the corresponding  variational free energy
yields a force 
$$\eqalign{
F \sim  (f^2\lb/b)^{1/3} \ln(F N^{1/2})^{1/3}
}\eqno(F)
$$
and a chain size
$$\eqalign{
R \sim N  (f^2\lb/b)^{1/3} \left[ \ln\left(N(f^2\lb/b)^{2/3}\right)\right]^{1/3}
}\eqno(R2)
$$
The results for the chain size and chain structure factor
were shown to agree {\it quantitatively}
with Monte-Carlo calculations \refto{BB93}.
This model can also be used to give a quantitative meaning to the 
"electrostatic blob size" introduced above. This is done by noting that
in the chain under tension, the mean squared distance between
two monomers separated by $n$ monomers along the chain is 
$R(n)^2= nb^2 + n^2 (b^2 F/3k_BT)^2$. The blob size can be obtained by identifying the second
term
with $((n/g)\xi_e)^2$,  $\xi_e=3k_BT/F$, in agremeent
with (\call{blob}). The blob size 
is therefore directly related to the prefactor of $n^2$ in 
the expression of $R(n)^2$.

In the gaussian variational approach, the trial Hamiltonian is 
a general quadratic form of all monomers coordinates. This method, which is 
tractable for cyclic chains only, was first introduced
by des Cloizeaux \refto{DJ85} as an alternative to Flory's theory
for chains with short range excluded volume. In that case, however, the
method gives a swelling exponent $\nu=2/3$ in three dimensions, and
is therefore inferior to Flory's approach. For long range interactions. It
was shown \refto{BP92} that the variational gaussian approach 
predicts the same asymptotic behaviour 
 as Flory's theory and as the "chain under tension"
method. As for the chain under tension model, the quantitative
agreement with  numerical simulation results is good \refto{BB93}.

\smallskip

{\it 2.3 Renormalisation group calculations}

The coincidence of the predictions from the Flory approach
and the variational gaussian method suggests that the predicted
asymptotic behaviour is indeed exact, except may be for logarithmic 
corrections. Exact renormalisation group calculations
were performed for space dimensions $d>4$ \refto{PV77,GE77}, showing that 
for  $4<d<6$ the swelling exponent of the chain is
$\nu=2/(d-2)$ (as does the variational calculation of 
reference \refto{BP92}). No such calculation 
is available for $d=3$, but the result for $d=4$ suggests that the maximum
value $\nu=1$ is obtained for $d<4$, while gaussian behaviour
is recovered for $d>6$. Moreover, 
a "naive" real space renormalisation
argument \refto{GE79} also predicts $\nu=1$. Note that a naive extension of the Flory 
theory to $d$ dimensions predicts an upper critical dimension where the chains are 
gaussian equal to $6$ but a lower critical dimension where $\nu=1$ equal to $3$ 
in disagreement with the renormalisation group results.

The success of variational and mean-field methods for chains with long range
interactions is not unexpected. As pointed out in \refto{BP92}, these
methods usually fail because a poor approximation 
of the monomer-monomer correlation function is used in the calculation
of the potential energy. A gaussian correlation function will
be maximum when two monomers are in contact, whereas the true correlation
function vanishes in that case.  For long range forces, 
 most of the potential energy comes from configurations
in which the monomers are far apart, so that the gaussian approximation becomes much better.

\smallskip

{\it 2.4 Screening of  electrostatic interactions}

When a finite concentration of salt is present in  the 
solution - which is always the case in experiments, at least because of water dissociation-, 
the solution becomes a conducting, rather than dielectric, medium. 
The fundamental implication is that  the  Coulomb interaction between charged 
monomers is screened by the salt solution, i.e. the electrostatic
potential created by a monomer, or a group of monomers, falls
 off exponentially rather than algebraically with distance. Thus 
distant parts of the chain do not interact, and the chain can
be expected to behave, in the asymptotic limit $N\rightarrow \infty$, as
a random walk with short-ranged repulsive interactions.

It is a general result from the theory
of charged solutions that charge fluctuations become
uncorrelated (or, equivalently, that an external charge distribution is 
screened) over a typical distance $\kappa_s^{-1}$, where $\kappa_s$
is related to the thermodynamic properties of the solution 
\refto{HM86}.
 The screened
interaction between 
monomers, mediated by the salt solution, can formally be obtained by 
integrating out the degrees of freedom associated to the salt ions 
in the partition function as shown in Appendix A. The resulting interaction is in 
general a complex function of the monomer coordinates, involving
many body interactions. In the case of dilute salt solutions
and sufficiently weak perturbations, however, the linear
response (or Debye-H\"uckel) theory can be used, and the 
 salt can be treated as an ideal gas. The screening length $\kappa^{-1}$
is then given by 
$$\eqalign{
\kappa^2 = 4 \pi\lb I,
}\eqno(kappa)
$$
 where $I$ is the ionic strength of the solution defined as
$I=\sum_{\rm ionic\  species} Z_i^2 c_i$, $Z_i$ being the 
valence and $c_i$ the concentration of species $i$.

The effective interaction can in this limit be 
written as a sum of pairwise additive
 interactions between the monomers. The effective pair potential 
 is given by the Debye-H\"uckel formula \refto{IS85}
$$\eqalign{
v_{DH}(r)=k_B T {\lb \over r } \exp(-\kappa r)\ .
}\eqno(vdh)
$$

This short range interaction potential is the starting
point of many theoretical studies of polyelectrolyte solutions. 
The definition of an effective pair potential requires 
however two conditions. The first condition,
  that the ionic
solution is dilute, is usually satisfied in polyelectrolyte solutions. The 
second condition of a weak external perturbation 
introduced by the polymer on the small ion solution must be 
considered more carefully, and is addressed in the next section.

Although the potential (\call{vdh}) is short-ranged, its range $\kappa^{-1}$
can be much larger than the monomer size, which is the typical interaction
range in neutral polymers. Depending on the ionic strength, 
$\kappa^{-1}$ can vary typically from less than $1nm$ to more than $100nm$. At distances 
smaller than $\kappa^{-1}$, the screened potential (\call{vdh}) is close to the unscreened 
Coulomb potential discussed in section (2.1).  The
behaviour and properties of polyelectrolyte chains can therefore
be markedly different from those of neutral chains over rather 
large length scales. In fact, section 4 will show that 
in some cases the effect of electrostatic interactions extends over
length scales much larger than $\kappa^{-1}$.

Finally, it should be noted that the concept of effective
 interaction between monomers, obtained by integrating out 
the degrees of freedom of the small ions, is useful only
 for the calculation of {\it static} properties. 
Dynamical properties such as diffusion constants or viscosities do not have
to be identical to those of an hypothetical polymer solution in which the
monomer-monomer interaction would be given by (\call{vdh}). The ionic
degrees of freedom can play a nontrivial role in the dynamics,
as they do in suspensions of charged collo\"{\i}ds \refto{RS89}.

\smallskip

{\it 2.5 Annealed and quenched polyelectrolytes}

An important degree of freedom of polyelectrolyte chains is the distribution of the charges 
along the chemical sequence. This distribution can be either quenched or annealed \refto{annealed}. 
When a weakly charged polyelectrolyte is obtained  by copolymerisation of a 
neutral and of a charged monomer, the number of charges on each chain and the positions 
of the charges along the chain are fixed, the distribution is quenched. 
If the charges are sufficiently regularly distributed along the chain and 
do not have a tendency to form long blocks, the precise distribution 
of the charges does not seem to be a relevant variable for the chain statistics. 
At the level of scaling laws that we use, one can expect that only the numerical 
prefactors are different when polyelectrolytes with a random  
or a periodic distribution of charges are compared.

Polyacids or polybases are polymers where 
the monomers can dissociate depending on the pH of the solution and acquire a charge. 
The dissociation of an $H^+$ ion from a polyacid gives rise to the apparition of a 
$COO^-$ group and thus of a negative charge. This is an annealed process, the total number of 
charges on a given chain is not fixed but the chemical potential of the $H^+$ ions and thus the 
chemical potential of the charges is imposed (by the pH of the solution). The positions of 
the charges along the chain are also not fixed; the charges can move by 
recombination and redissociation of an $H^+$. The chemical potential of the charges $\mu$ 
depends on the fraction $f $ of dissociated acid groups and
is related to the pH of the solution by \refto{Mandel}
$$\eqalign{
pH=pK_0 + \mu(f).
}\eqno(ann)
$$
The charge chemical potential has two contributions, an entropic 
contribution related to the mixing of charged and non charged groups along the chains 
$k_B T \ln [f/(1-f)]$ and an electrostatic contribution 
$\mu_{el}(f)= N^{-1} {\partial F_{el}/ \partial f}$ which 
is the derivative of the electrostatic free energy of the chain. In a good or $\theta$ solvent, 
the electrostatic contribution $\mu_{el}$ is an increasing monotonic function of $f$ 
and for a given pH the fraction of charges $f$ is well defined, the properties of the chains 
are very similar to those of a quenched chain with a charge fraction $f$. In a poor solvent, 
the blob model introduced by Khokhlov 
and discussed above leads to a non monotonic variation of $\mu_{el}$ and thus of $\mu$ with $f$. 
This non monotonic variation indicates a conformational transition of the chain 
between a collapsed weakly charged conformation and a stretched strongly charged conformation. 
The collapse transition of a polyelectrolyte with varying $f$ is thus 
predicted to be a first order transition \refto{annealed}. 
The discontinuous 
collapse of the chains could explain titration curves of polyacids or 
polybases in a poor solvent (polymethacrylicacid for example) which show a plateau in the 
$pH-f$ curve \refto{Petitpas}.

\bigskip

{\bf 3. Local aspects of screening}

\smallskip

\taghead{3.}

The Flory-like model presented in the previous section 
considers only the interactions between
charged monomers along the polyelectrolyte chain and 
ignores the role of the small counterions 
that neutralize the polyelectrolyte charge. When 
the polyelectrolyte is strongly charged, the electrostatic 
potential on the chain is large and some of the counterions remain bound
to the chain. 
This phenomenon is known as counterion condensation, or Manning condensation
\refto{MA68}.
For many purposes 
one must then consider that the chain has 
an effective charge due to the charged monomers 
and to the sheath of bound, or condensed, counterions.
 The effective charge is lower than the 
nominal charge of the monomers. Counterion condensation can be 
effective even at infinite dilution and reduces the electrostatic
 interactions between monomers.
We first give here the more qualitative description of 
the condensation due to Manning \refto{MA68} and then 
present an earlier discussion based on the
 Poisson-Boltzmann equation due to Fuoss et al. \refto{FK}. 
The condensed counterion sheath around a 
polyelectrolyte chain is electrically polarizable;
 this can induce 
attractive interactions between polyelectrolyte chains.
 We discuss in section 3.2. possible  mechanisms for
attractive electrostatic interactions in polyelectrolyte solutions.  

\smallskip

{\it 3.1 Counterion condensation}

For simplicity, we consider a negatively 
charged polyelectrolyte chain, in a salt-free solution, which has 
locally a rodlike conformation and  can
 thus be considered as a rod over 
a distance $L$  much larger than 
the monomer size  $b$. The distance between charges 
along the chain is A and
the fraction of charged monomers is $f=b/A$. If we
 ignore the contribution of the counterions,
the electrostatic potential at a distance 
$r<<L$ from the chain is obtained from Gauss theorem as $\psi (r)={2\lb /A} \ln(r)$ 
(we use here as unit of 
the electrostatic potential $k_BT/e$ where $e$ is the elementary charge). 
The distribution of the counterions around the chain is given by Boltzmann
statistics, for monovalent counterions,  
$n(r)=n_0 \exp-{\psi (r)}\simeq r^{-(2\lb /A)}$ where $n_0$ is a constant. 
The total number of counterions 
per unit length within a distance $r$ from the chain is then 
$$\eqalign{
p(r)=\int\limits_{0}^r 2\pi r'dr'n(r')=\int\limits_{0}^r 2\pi n_0 r'^{(1-2\lb /A)}dr' 
}\eqno(man1)
$$
When the charge parameter $u=\lb /A$ is smaller than one, 
the integral giving $p(r)$ 
is dominated by its upper bound and $p(r)$ decreases 
to zero as the distance $r$ gets small. 
In this case where the  charge fraction is small
 $f<b/\lb$ there is no counterion condensation 
on the polyelectrolyte chain.
In the opposite limite where the charge is
 large $u=\lb /A>1$, the integral giving p(r) 
diverges at its lower bound indicating a strong 
condensation of the counterions on the 
polyelectrolyte chains. As the condensation proceeds,
 the effective value of the charge parameter 
decreases. 
The counterion condensation stops when the effective
 charge parameter is equal to one. 
The polyelectrolyte chain 
and the condensed counterion sheath are then 
equivalent to a polyelectrolyte with a 
distance between charges $A=\lb$ or to 
a fraction of charged monomers $f=b/\lb$. 
The remaining monomeric 
charges are neutralized by the condensed counterions.

 One of the directly measurable quantities that
 strongly depends on the condensation 
of the counterions is the osmotic pressure 
of the polyelectrolyte solution. 
As shown below, in many cases the osmotic 
pressure of a polyelectrolyte solution 
is dominated by the counterions,
 the contribution of the polyelectrolyte 
chains being only a very small correction.
 Below the Manning condensation threshold, 
the counterions are not condensed and are 
essentially free. At leading order, 
the solution can be considered as an ideal
 gas of counterions of concentration $fc$ 
where $c$ is the total monomer concentration 
and the osmotic pressure is \refto{JP80}
$$\eqalign{
\pi= k_B T fc
}\eqno(manp)
$$   
The first correction to the ideal gas behavior is due to the polarisation 
of the counterion gas by the polyelectrolyte chains. 
The polarisation energy is 
estimated using the Debye-H\"uckel approximation for the
 gas of counterions \refto{LL}. The screening 
is due to the counterions and 
$\kappa^2= 4\pi\lb fc$. The polarisation free 
energy per unit volume reads
$$\eqalign{
F_{pol}= -k_B T fcu \ln (\kappa) = -{k_B T  \over 2} c ({b\lb \over A^2})\ln (4\pi\lb c b/A)
}\eqno(manpol)
$$
The osmotic pressure of the solution is directly calculated from this relation
$$\eqalign{
\pi= k_B T fc(1-u/2)
}\eqno(manp')
$$
In many cases, for weakly charged polyelectrolytes, $\lb\simeq b$,
so that $u \simeq f$,
and the polarisation correction is small. 
In most of the following, we  use the ideal
gas expression for the osmotic pressure of the counterions in a weakly 
charged polyelectrolyte solution.

Above  Manning condensation threshold $u>1$, 
the bound counterions do not contribute to the osmotic pressure.
The remaining counterions, at a 
concentration $c b/\lb=fc/u$, behave as a Debye-H\"uckel
gas, polarized by a chain which has an effective
charge parameter equal to unity.
 The osmotic pressure
 is given by equation (\call{manp'}), where $fc$ is replaced by 
$c b/\lb$ and $u$ by $1$, 
$$\eqalign{
\pi= k_B T cb/(2\lb) = k_B T fc/(2u) \ .
}\eqno(manpres)
$$

Several other physical quantities such as the electrical 
conductivity of the solution or 
the electrophoretic mobility of the chains \refto{mobility} 
 strongly depend
 on the condensation of 
the counterions and are dominated by the free counterions. 
Their determination is, in general, in good agreement with 
 Manning condensation theory.

\smallskip

{\it 3.2 Poisson-Boltzmann approach}

Fuoss, Katchalsky, and Lifson  \refto{FK} have studied the interaction between 
the polyelectrolyte chains and their counterion clouds using 
the so-called Poisson-Boltzmann equation. They consider a solution 
of infinite rodlike molecules, which are all parallel.
 The number of molecules per 
unit area is $\Gamma$. Around each of these molecules, there
 exists an equipotential surface where 
the electric field vanishes. On average, this surface can
 be approximated by a cylinder parallel 
to the polyelectrolyte chains with a radius 
$R$ defined such that $\Gamma \pi R^2=1$. The average
density of counterions in the cylinder is $c_i=fc=1/(\pi R^2A)$ 
where $A$ is the distance
 between charges along the chain. 
>From an electrostatic point of view, each chain with its 
counterions in the cylinder of radius $R$ is independent 
of all the other chains. 
The electrostatic problem 
that must be solved is therefore that of a single infinite chain with its 
counterions embedded in a cylinder of radius $R$ with 
the boundary condition that the electric field vanishes 
on the surface of the cylinder.

The electrostatic potential in this cell satisfies the Poisson 
equation $ \Delta \psi= -4\pi\lb c_i(r)$
where $c_i(r)$ is the local
 counterion concentration at a distance $r$ from the chain. 
The local counterion concentration is obtained 
from Boltzmann statistics $c_i(r)=c_i\exp\left(-\psi(r)\right)$. 
This leads to the Poisson Boltzmann equation
$$\eqalign{
 \Delta \psi(r)=-\kappa^2 \exp\left(-\psi(r)\right)
}\eqno(pbeq)
$$
The Debye screening length is defined here
 as $\kappa^2= 4\pi\lb c_i$; it is such 
that $\kappa^2R^2=4u$ where $u$ is the 
charge parameter defined above. 
The Poisson Boltzmann equation 
has been solved exactly in this geometry in reference \refto{FK}. 
The expression of the potential critically
depends on the value of the charge parameter.
 Below the Manning condensation 
threshold where $u<1$
the electrostatic potential is given by 
$$\eqalign{
\psi=\ln [{\kappa^2r^2\over 2\beta^2} \sinh^2(-{\rm arctanh}\beta + \beta \ln (r/R)]
}\eqno(pot1)
$$
where $\beta$ is an integration constant related to $u$ by 
$u =(1-\beta^2)(1-\beta / {\rm tanh}(\beta\ln(b/R))^{-1}$

Above the Manning condensation threshold the electrostatic potential reads
$$\eqalign{
\psi=\ln \left[{\kappa^2r^2\over 2\beta^2} \sin^2(-{\rm arctan}\beta + \beta \ln (r/R)\right]
}\eqno(pot2)
$$
and $\beta$ is given by $u =(1-\beta^2)(1-\beta / \tan(\beta\ln(b/R))^{-1}$. 
The distribution of counterions around each polyion can then be determined 
using Boltzmann's statistics. 

At each point of the solution, the pressure has two contributions, an ideal gas 
contribution and an electrostatic contribution proportional 
to the square of the local electric field. At the edge of the cell ($r=R$), 
the electric field vanishes and there is only an ideal gas contribution.  
The pressure being uniform throughout the solution can be calculated at this point
$\pi=k_B T c_i(r=R) =k_B T c_i \exp-\left(\psi(r=R)\right)$. 
Both above the Manning condensation thereshold (\call{manp'}) and below 
the threshold (\call{manpres})
the calculated pressure is in agreement with that 
calculated from Manning theory.

Manning theory describes the condensation of the counterions as 
a transition between two states, a bound state and a free
state. In the Poisson Boltzmann approach, 
the distribution of the counterions 
is continuous and there is no bound state. However,
 close to the polyelectrolyte chain, 
the interaction energy between one                    
 counterion and the polyelectrolyte chain becomes 
much larger than $ k_B T$  (in the region where 
$\psi>>1$). One could then divide space into two regions, 
a region close to the chain where the 
interaction energy is larger than $k_B T $
 (with a size of the order of a few times the monomer size $b$)
and where the counterions can be considered as bound and a region  further away 
from the chain where the interaction between a counterion and 
the chain is smaller than $k_B T$
  and where the counterions can be considered as unbound.

{\it 3.3 Attractive electrostatic interactions}

The counterions condensed on a polyelectrolyte chain 
 move essentially freely along the chain; they  more or less form
a one dimensional gas of average  density $g/b$ where $g=f-b/\lb$ is 
the fraction of condensed counterions. The charge density along 
the chain is therefore not a frozen variable and shows thermal fluctuations due 
to the mobility of the counterions. When two chains are sufficiently close, 
the charge density fluctuations on the two chains are coupled by the 
electrostatic interactions; this leads to attractions between 
the polymers which are very similar in nature to the Van der Waals 
interactions between polarizable molecules. 
A fluctuation induced attraction can be 
expected for all polyelectrolytes where the charges are mobile along the chain. 
This is for example the case above Manning condensation threshold but also
for annealed polyelectrolytes, which are polyacids or polybases, where 
the charge can be monitored by tuning the pH of the solution \refto{ annealed ?}. 
The ionised groups 
are not fixed on these chains, only the chemical potential of 
the charges is imposed and the charge density also has thermal fluctuations.

We first calculate the attractive interaction for 
two parallel rodlike molecules of length $L$  at a distance $x$
in a solution where the salt concentration is $n$, 
following the ideas introduced by Oosawa \refto{OS71}. 
When a charge fluctuation $\delta c_1(z)$ 
occurs along the first chain it polarizes the second chain where a
 charge fluctuation of opposite sign
$\delta c_2(z)$ appears.
 The attraction between the two chains 
is due to the Coulomb interaction between these charge fluctuations.
Within a linear response approximation,
the free energy of the charge fluctuations can be written in Fourier space as 
 $$\eqalign{
H={k_B T \over 2} \sum_{q} \sum_{i,j}S_{ij}^{-1}\delta c_i(q)\delta c_j(-q)
}\eqno(att1)
$$
$\delta c_i(q)$ is the Fourier transform at a wave vector q of 
the charge fluctuation  $\delta c_i(z)$ ($i=1,2$).
 The charge structure factor on one of the chains 
is $S_{11}(q)=g/b$ and corresponds to a one-dimensional
 ideal gas of charges. The crossed structure 
factor is due to the electrostatic interactions between the two chains 
obtained by summing the Debye-H\"uckel interactions
$$\eqalign{
S_{12}^{-1}(q)=4\pi \lb \int{{d^2k\over {4\pi^2}}{e^{iqx\cos\theta} \over {k^2+q^2+\kappa^2}}}
=2\lb K_0[x(q^2+\kappa^2)^{1/2}]
}\eqno(att2)
$$
where $K_0[x]$ is a modified Bessel function  of the second kind \refto{AS}.
The fluctuation contribution to the interaction free energy
between the two rods 
is obtained by summing the partition function
 $\exp-H/k_B T $ over all fluctuations. We obtain
$$\eqalign{
F_{att}(x)=L{k_B T \over 2}\int {dq\over {2\pi}}\ \ln\left(1-4{g^2\lb^2\over 
{b^2}}K_0[x(q^2+\kappa^2)^{1/2}]\right)
}\eqno(att3)
$$
This expression can be simplified by assuming that the fraction $g$ of recondensed 
counterions is smaller than one and by looking for the asymptotic 
limits where the distance $x$ between 
the rods is smaller or larger than the screening length $\kappa^{-1}$.
 For short distances between the rods, the interaction energy is
$$\eqalign{
F_{att}(x)=-\alpha k_B T {g^2\lb^2\over {b^2}}{L\over x}
}\eqno(att4)
$$
where $\alpha$ is a numerical constant. For large distances,
 the attractive interaction decays as
$$\eqalign{
F_{att}(x)=-\beta k_B T {g^2\lb^2\over {b^2}}{L \over{x^{3/2}\kappa^{1/2}}}\exp(-2 \kappa x)
}\eqno(att5)
$$
$\beta$ being another numerical constant. As expected the interaction is attractive. 
It is proportional to the square of the
Bjerrum length indicating that this is a second order electrostatic effect: 
the fluctuation on the first rod creates an electric
 field that polarises the second rod; 
the charge fluctuation on the second rod then creates an electric field which 
interacts with the charge fluctuation of the first rod. 
For parallel rods, the interaction 
is also proportional to the length of the rod, each 
segment interacts mainly with the segment 
of the other rod which is directly facing it. For the same reason,
the attractive interaction is screened over a length $\kappa^{-1}/2$, i.e. over
half the screening length of the repulsive coulombic interaction. 

When the two rods make an angle $\theta \ (0<\theta<\pi/2)$, 
the attractive electrostatic interaction cannot be calculated 
from the fluctuation free energy (\call{att2}). Instead, we directly discuss 
it in terms of the polarisability of the two polyelectrolytes 
\note{ This is equivalent 
to performing a perturbation expansion to second order
in the crossed term that appears in equation (\call{att2}).}
If a charge fluctuation
$\delta c_1(q)$  is created on the first rod, 
it creates an electrostatic potential on the second rod
 $\delta \psi (q') = K(q,q')\delta c_1(q)$ where
the kernel is defined as 
$K(q,q')={2\pi \lb \over {\sin \theta}}{\exp (-px) \over p}$  
with a wavevector 
p related to $q$ and $q'$ by 
$p^2= \kappa^2+{{q^2+q'^2-2qq'cos\theta} \over {\sin^2 \theta}}$. 
The response function of the one dimensional gas of recondensed 
counterions on the second rod is $g\over b$ and the 
induced charge fluctuation on the second rod is
$\delta c_2(q')=- {g\over b}K(q,q')\delta c_1(q)$. 
This charge fluctuation creates a potential on the 
first rod $\delta \psi'(q)=K(q,q')\delta c_2(q')$. 
The attractive interaction energy between the two 
rods forming an angle $\theta$ at a distance $x$
is then estimated as
$$\eqalign{
F_{att}(x)=-k_B T  {g\over b}\int{{dqdq'\over {4\pi^2}} K^2(q,q') <\delta c_1^2(q)>}
}\eqno(att6)
$$
The average value of the charge 
fluctuation is given by the ideal gas statistics $<\delta c_1^2(q)>=g/b$.
 This leads to an attractive interaction between non parallel rods 
$$\eqalign{
F_{att}(x)=-k_B T  {{g^2\lb^2}\over {b^2 \sin \theta}} (\pi -\theta)
  {\rm E_i}(2\kappa x)
}\eqno(att7)
$$
where ${\rm Ei}(u)$ is the exponential integral function \refto{AS}. As for
 parallel rods,
 the attractive interaction is a second order electrostatic effect, 
it is proportional to $\lb^2$ and it decays at
 large distances as $\exp{-2\kappa x}$. 
When the angle between the two rods is not very small, 
the attractive force is independent of the length of the rods. 
In this geometry only the chain segments
 in the vicinity of the crossing point interact.
If the distance between the rods
is smaller than the screening length, the crossover to the parallel 
rod behavior occurs when the angle $\theta$ is smaller than $x/L$.
At short distances $x<\kappa^{-1}$, ${\rm Ei}(u)=\ln u$ and the attractive 
interaction energy varies logarithmically. 
As the attraction is independent of the chain length, 
the attractive interaction between two polymers which are rodlike 
over a distance larger than the screening length is also given by equation (\call{att6}).

The fluctuation attractive interaction must be compared to the repulsive 
coulombic interaction between the chains. For two rods at
a distance x that make a finite angle, 
the repulsive Coulomb interaction is of order
 $Fel \simeq k_B T  2\pi \lb f^2/(\kappa b^2) \exp{-\kappa x}$ where 
$f$ is the effective fraction of charged monomers 
(equal to $b/\lb$ above the condensation threshold). 
At large distances, as already mentionned,
the screening of the attractive interaction is stronger 
and the repulsive interaction is dominant. At small distances 
the attractive interaction is dominant. The distance at 
which the attractive force is larger than the repulsive 
force is of order $x\simeq \lb (g/f)^2$. In general this corresponds 
to very short distances of the order of the Bjerrum length.
The attractive fluctuation 
interaction is thus expected to be important only in 
strong coupling situations where the Bjerrum length is large 
(for example in the presence of multivalent ions) 
or if some external constraint imposes very 
small distances between chains. 

The attractive interaction can also
 be considered within the framework of the Poisson-Boltzmann 
approach. The Poisson-Boltzmann equation
 is derived from a mean field theory that considers only 
the average concentration profile of the counterions around 
each polyelectrolyte chain. A more refined theory should take 
into account the thermal fluctuations around this concentration profile. 
Such a 
theory has been built to study the interactions between 
charged colloidal particles and also leads to 
attractive interactions \refto{Marcelja,Robbins}.
 Qualitatively, the conclusions 
are the same as the one presented here, 
the attractive interactions are dominant at short distances 
and becomes relevant only in the limit of strong electrostatic coupling.

Experimentally, the addition of multivalent ions to polyelectrolyte
solutions often provokes a precipitation. This can be explained 
by attractive electrostatic interactions \refto{Belloni} or by complexation
of the polymer by the multivalent ions \refto{Wittmer}.

The interaction between charge fluctuations is not 
the only mechanism that leads to attractive
 interactions between polyelectrolyte chains. Recently, 
Ray and Manning \refto{MR94} have presented a model for attractive
 interactions based on the overlap between 
the condensed counterions sheaths
around the two interacting polymers. In the simple version of Manning 
condensation theory presented here, the polyelectrolyte
 chains are infinitely thin lines
and the pointlike counterions condense directly  
on the line. A more refined version
allows for a finite condensation volume that 
can be determined by minimization of the free energy
of the condensed counterions. When two parallel rods come close to one another, 
the condensation volume around each rod expands 
in the space between the two rods and 
the two condensation volumes overlap. This leads to the
formation of a polyelectrolyte dimer with 
a single condensation sheath. The expansion of the condensation 
volume leads to an increase in 
translational entropy of the counterions and thus to
 an attractive interaction. The model 
shares thus some analogy with the 
capillary condensation between thin films wetting parallel cylinders 
\refto{Yeomans}. An attractive interaction is predicted at 
intermediate distances much larger than the polyelectrolyte 
diameter but much smaller than the electrostatic screening length.
 This interaction has been 
calculated only  for parallel polyelectrolyte rods, 
it has not been calculated for rods intersecting at a finite angle.   

\bigskip

{\bf 4. Electrostatic rigidity}

\taghead{4.}

\smallskip

In this section, we assume that the interaction between charged monomers
can be described by the Debye-H\"uckel potential, (\call{vdh}). 
Within this approximation, we discuss
the structure of the polymer chain at length scales intermediate between
the short length scales discussed in section 3. and 
the chain size. It has already been mentioned in section 2.4 that at very 
large scales, a charged chain has the same structure as a neutral polymer
chain in a good solvent, since the potential (\call{vdh}) is short ranged. 
The radius increases with molecular weight as $R\simeq N^{3/5}$ (in the Flory approximation).
At short scales ($L<\kappa^{-1}$), on the other hand, screening
is inoperant, so that the chain is expected to take the rodlike
structure described in section 1. The simplest assumption 
concerning the chain structure at intermediate length scales,
which was made in all early work in the field \refto{PF78}, 
is that
the crossover from rodlike to Flory-like behaviour
takes place above a length scale $\kappa^{-1}$, equal to the range
of the interactions. This simple assumption was challenged
by the works of Odijk \refto{OD77,OH78} and Skolnick and Fixman
\refto{SF77}, who showed how the Debye-H\"uckel
 interaction can induce a rodlike conformation at length scales 
much larger that the  interaction range.
 Their theory, and its limitations,
are presented here.

\smallskip

{\it 4.1 The Odijk-Skolnick-Fixman theory}

In their calculations,  Odijk-Skolnick and Fixman (OSF) consider a
semi-flexible chain with total contour length L characterized
by its "bare" persistence length $\l0$, and that carries charges
separated by a distance $A$ along its contour. The interaction 
between the charged monomers is given by (\call{vdh}), and only 
electrostatic interactions are taken into account
(the polymer backbone is in a theta solvent). For strongly charged
chains with $A  < \lb$ Manning condensation 
can in a first approximation be accounted for by
replacing $A$ with  $\lb$. In the limit $\l0<< A $ of a
weakly charged chain, this model is equivalent to the gaussian chain
model discussed in section (2.1), with $N=L/A$, $b^2= A \l0$ and $f=1$.
We consider here the general case, where $\l0/A$ is not small \refto{BJ93}.

The total energy of the chain is the sum of the intrinsic curvature energy
and of the (screened) electrostatic energy. If the polymer is described 
 as a planar curve  of  curvilinear length $L$, a 
chain configuration is specified by the function $\theta(s)$ defined
for  $- L/2 < s <  L/2$ by
$\cos(\theta(s)) =  \ttt (s) \cdot \ttt (0) $
where  $\ttt(s)$ is the unit vector tangent to the chain at
the point of curvilinear abscissa $s$. Assuming that $\theta(s)$ remains small,
the electrostatic potential
between two charges located at $s_1$ and $s_2$ on the chain may be expanded
around
the value obtained for a rodlike configuration,
$v_{DH}(|s_2-s_1|)$. The following expression is then
obtained for the
total energy  $H[\theta]$ 
 of a given configuration  $\{\theta(s)\}$:
$$\eqalign{
H[\theta] = H_0 + {1 \over 2} k_B T \int_{-L/2}^{L/2}ds \ \int_{-L/2}^{L/2}ds'
\  {d\theta(s) \over ds} \left[ \l0 \delta(s-s') + K(s,s')\right]
{d\theta(s') \over ds}\ .
}\eqno(en)
$$
Here $H_0$ is the electrostatic energy of a rod,
the term proportional to $\l0$
 is the "bare" curvature energy of a noninteracting
semiflexible chain, and the contribution
of the electrostatic interactions is described by the kernel $K(s,s')$
which reads (for $s>s'$):
$$\eqalign{
K(s,s')= {1 \over A^2} \int_{-L/2}^{s'}ds_1\ \int_{s}^{L/2} ds_2
\ v_{DH}'(s_2-s_1) {(s_2-s) (s'-s_1)\over (s_2-s_1)}
}\eqno(K)
$$
with $v_{DH}'(s)={dv_{DH} \over ds}$
The only assumption made in obtaining this result
is that, within the range $\kappa^{-1}$ of the interaction potential,
the chain remains in an almost rodlike configuration, i.e.
$\vert s_2-s_1\vert - \vert {\bf R}(s_1) - 
{\bf R} \vert  << \vert s_2-s_1\vert $
for $\vert s_2-s_1 \vert < \kappa^{-1}$. For long chains, 
the integration in (\call{K}) can be extended
to infinity, and $K(s,s')$ becomes a function
of $s-s'$,
$$\eqalign{
K(s) =  {1\over 6 A^2} \int_0^\infty dx\ {x^3 \over x+s}
v_{DH}'(x+s)  \ .
}\eqno(K1)
$$
The statistical properties of the chain are obtained by 
integrating the Boltzmann factor $\exp(-H[\theta]/k_BT)$ over all 
possible configurations, i.e. over all functions $\theta(s)$. The 
integration can be carried out analytically, since the energy is a quadratic 
function of $\theta$. The calculation is simplified by introducing
the Fourier transform of the kernel (\call{K1}),
$$\eqalign{
\til{K}(q) & = \int_0^{\infty} ds \exp(i q s) K(s) \cr
     & =  \lo {2 \kappa^2  \over  q^2} \left(  {\kappa^2+q^2 \over q^2}
 \ln\left({\kappa^2+q^2 \over \kappa^2}\right)  
-1  \right)\ , 
}\eqno(Kq)
$$
where  
$$\eqalign{
\lo =\lb / (4 A^2 \kappa^2)
}\eqno(lpe)
$$ 
has the dimension of a length, and
is known as  the Odijk-Skolnick-Fixman length.
The local flexibility of the chain can be characterized by 
the mean squared angle $<\theta(s)^2>$ between the 
chain direction at the origin and after a contour length $s$. For a 
neutral semiflexible chain,  $<\theta(s)^2> = s/\l0$ varies linearly with $s$.
The chain configuration can be described as resulting from an "angular
diffusion" process, with a diffusion constant $\l0^{-1}$.
For the charged chain, with the approximate energy (\call{en}), one gets
$$\eqalign{
<\theta(s)^2 >= {4\over \pi}
 \int_0^\infty
 dq\  { {\rm sin}^2(qs/2)  \over q^2} \ {1 \over
\l0 + \til{K}(q) }\ \ ,
}\eqno(th2)
$$
This expression simplifies in the limits of small $s$ and large $s$. 
For large $s$, 
$$\eqalign{
<\theta(s)^2> = {s \over \l0+\lo} 
}\eqno(odijk)
$$
which expresses the fact that at large scales, the chain conformation can 
be described by an effective persistence length 
$\l0+\lo$, which is the sum of a "bare" 
and of an electrostatic contribution. This is the well known result 
first obtained in references \refto{OD77} and \refto{SF77}. It indicates
that the influence of the screened electrostatic interactions 
can extend much beyond their range $\kappa^{-1}$, since $\lo$ is,
for weakly screened solutions, much larger than  $\kappa^{-1}$. The persistence length 
also decreases with the salt concentration as $\lo \simeq n^{-1}$ whereas the Debye 
screening length has a slower decay $\kappa^{-1} \simeq n^{-1/2}$.

For small values of $s$, (\call{th2}) reduces to 
$$\eqalign{
<\theta(s)^2> = {s \over \l0},
}\eqno(bare)
$$
The chain statistics at short scales are not modified
 by electrostatic interactions. The crossover between the "intrinsic" regime
described by (\call{bare}) and the "electrostatic" regime described
by (\call{odijk}) takes place when the electrostatic interactions become
strong enough to perturb the statistics of the neutral flexible chain. The 
crossover length $s_c$ can be obtained qualitatively from the following argument
\note{ From its definition, $s_c$ can be described as the 
contour length of an "orientational blob"}.
 If a small chain section, of
 length $s < \kappa^{-1}$, is bent to form an angle $\theta$, the cost 
in "bare" curvature energy is $k_B T \l0 \theta^2 /s$, while the cost
in electrostatic energy is $ k_B T \lb  (s/A)^2 (\theta^2/s) $. The two 
energies are comparable for $s\simeq s_c$, which gives
$s_c \sim A (\l0/\lb)^{1/2}$. 

A more detailed treatment of the crossover regime is
 possible by ignoring logarithmic factors and 
replacing the exact Kernel (\call{Kq})
by the approximate expression, $\til{K}(q) = \lo \kappa^2/(\kappa^2+2 q^2)$.  
The integral 
(\call{th2}) can then be computed analytically, and yields,
$$\eqalign{
< \theta(s)^2 > = {s \over \lo+\l0} +
 {\lo \over \kappa \l0^{1/2} (\l0+\lo)^{3/2}} (1-\exp(-{s\over s_c}))
}\eqno(th21)
$$
with $s_c= \kappa^{-1} (\lo+\l0)^{-1/2} \l0^{1/2}$. This formula
crosses over from the "intrinsic" regime towards the electrostatic
regime for $s \sim s_c$. In the weak screening 
limit, $\l0<< \lo$ and the simple result $s_c\sim A (\l0/\lb)^{1/2}$
is recovered. 

The picture that emerges from this calculation is 
that the chain flexibility depends on the length scale. At short scales, 
$s<s_c$, the chain structure is determined by its  bare rigidity $\l0$,
 while the electrostatic rigidity (\call{lpe}) dominates at large scales.
Large scale properties, such as the giration radius or the structure factor
at small wavevector, can be determined by applying the standard formula
for semi-flexible chains of persistence length $\lo$. Excluded volume effects
between Kuhn segments of length $2\lo$ are accounted for by assigning
a diameter $\kappa^{-1}$ to each segment. The excluded volume between 
Kuhn segments is $\lo^2 \kappa^{-1}$ \refto{OH78}.

The only approximation required to obtain
the Odijk-Skolnick-Fixman length is the expansion that
yields equation (\call{en}). The calculation is therefore
consistent if 
the angle $<\theta (\kappa^{-1})^2>$ is small compared to unity.
It is easily checked that for $s\sim \kappa^{-1}$, the second
term dominates in the r.h.s. of (\call{th21}). The requirement 
$<\theta (\kappa^{-1})^2> << 1$ is then equivalent to 
$s_c \lo / (\l0 (\l0+\lo)) \simeq s_c/\l0 <<1$. i.e. the angular deflection
of the chain must be small when the crossover region is reached. 
In other words, the angular fluctuations 
that take place before the electrostatic interactions can come into play
and rigidify the chain 
should not be too strong.  If these fluctuations are too large, i.e.
if the chain is too flexible, the perturbation expansion that 
underlies the OSF calculation breaks down. The criterion for the validity of 
the calculation can be simply written, in the limit of weak screening,
 as $\l0 > A^2/\lb$. This implies that the OSF calculation should be directly
applicable to stiff chains such as DNA ($\l0 \sim 50{\rm nm}$), but has
to be reconsidered for flexible chains such as polystyrene sulfonate 
($\l0 \sim 1nm$).
Indeed, numerical simulations clearly confirm
 the behaviour described by equations 
(\call{Kq}) and (\call{th2}) for values of $s_c/\l0$ smaller than $0.2$
\refto{BJ93}. Deviations from this behaviour appear for larger values
of $s_c/\l0$.

Before closing this section, it is worth mentioning that 
 the electrostatic rigidity of a charged chain was 
computed numerically by Le Bret and Fiman \refto{LB82,FI82} using the 
Poisson-Boltzmann equation, rather than the Debye-H\"uckel 
approximation. At high ionic strength, the results deviate  significantly
from the OSF prediction, and the electrostatic rigidity
tends to behave as $\kappa^{-1}$ rather than $\kappa^{-2}$.

{\it 4.2 Alternative calculations for flexible chains}

The calculation presented in the previous section
shows that the Odijk-Skolnick-Fixman theory in its original form
is not applicable to flexible or weakly charged
 chains as such. A generalization of 
this theory was nevertheless used by Khokhlov and Katchaturian \refto{KK82}.
These authors propose that the expression (\call{lpe}) can be used 
for flexible chains, with the only modification that the distance 
between charges $A$ had to  be replaced by $\xi_e/(fg)$
 (see equation (\call{blob}). The bare persistence length
$\l0$ is also replaced by the blob size $\xi_e$
(A recent 
calculation of Li and Witten,
 including the thermal fluctuations of the chain gives
 a more quantitative basis to this approach \refto{LiWitten}). As a result,
the total persistence length of the chain reads
$$\eqalign{
\ell_{KK}= \xi_e + {1 \over 4 \kappa^2 \xi_e }\ .
}\eqno(lkk)
$$
The arguments of section (4.3) show, however, that the extension
of the OSF formula is rather speculative, since the condition that was
derived for the validity of the OSF theory is not met for a chain
characterized by the parameters  $\l0=\xi_e$ and $A=\xi_e/fg$. When the persistence 
 length is dominated by the electrostatic interactions, the radius of the chain 
is given by the Flory statistics
$$\eqalign{
R_{KK}= N^{3/5} \xi_e^{-4/5} \kappa^{-3/5} b^{6/5} \ .
}\eqno(Rkk)
$$
This generalisation of the Odijk theory is consistent if the Debye screening length is larger than
the electrostatic blob size $\xi_e$. At higher ionic strength, in the weak coupling 
limit there is no electrostatic rigidity and the polyelectrolyte
chain behaves as a flexible chain with 
short range interactions given by the Debye-H\"uckel formula.
 Its radius is given 
by equation  (4.12) below.

A number of approximate theories, based on  different variational ansatz,
have been proposed to describe the structure of flexible chains
\refto{SC91,BJ93, BD94, HT95}. 
These calculations differ at the technical level. The basic idea is to describe
a flexible charged chain (with screened interactions)
 by some model of noninteracting semiflexible chain, and 
to variationally optimize the persistence length of the noninteracting system.
All these variational calculations produce for flexible chains
 a persistence length that scales as $\kappa^{-1}$, i.e. the 
"naive" result of Katchalsky and Pfeuty. An exemple of such a variational 
calculation can be given for the particularly simple case of a 
freely jointed charged chain, with $\l0= A$. The variational ansatz used
to represent the system is that of a noninteracting 
chain in which the angle
between neighboring segments is bounded by a maximum value $\theta_{max}$, so
that the persistence length is $\ell_p= 2A/\theta_{max}^{-2}$. The variational 
entropy per segment  is then  $- \ln(\theta_{max})$. The 
variational energy can be expressed as a function of the structure factor 
of a chain with a persistence length $\ell_p$, $S_{ni}(\qq)$ as
$\int d^3{\qq} S_{ni}(\qq) v_{DH}(q) $. Using the expression of 
the structure factor obtained by desCloizeaux \refto{DC73}, the energy 
is the sum of the electrostatic energy of a rod and of a correction due
to the bending that can
be approximated by 
$ k_BT (\lb/A ) (1/\kappa \ell_p)$. Minimizing the sum of these two terms 
with respect to $\theta_{max}$, one finds $\ell_p \sim \kappa^{-1}$. This approach
can be straightforwardly generalized to the case where each rod in the freely
jointed chain represents an electrostatic blob \refto{BJ93}, in which
case it applies to the charged gaussian chain considered
in section 2.  The variational model is then 
that of a of a
 "semiflexible chain of electrostatic
blobs" , analogous to the "chain under tension" model (section 2), except
for the fact that the direction of the tension
is now fluctuating, and becomes uncorrelated over a persistence length  
 $\kappa^{-1}$.
 In that case, 
it must be realized that the "persistence length" $\kappa^{-1}$ is defined
in reference to the distance in space, rather than the contour length along 
the chain. The radius of the chain scales in this case as
$$\eqalign{
R_{P}= N^{3/5} \xi_e^{-3/5} \kappa^{-2/5} b^{6/5}\ .
}\eqno(RP)
$$

Variational calculations have the obvious drawback that 
the result crucially depends on the variational ansatz.
An ansatz too far from the actual structure
can yield incorrect results. Also, their range of validity is difficult
to assess. The variational calculation for gaussian chains
briefly described above is consistent only as long as 
the "electrostatic blob" concept can be used, 
This gives an upper limit for the charge density
of the chain for which its description by a
 "semiflexible chain of electrostatic
blobs" is possible, $ A^{-1} < (\l0\lb)^{-1/2}$ (this corresponds 
to $g>1$ in \call{blob}). This upper limit corresponds 
to the lower charge density for which the OSF calculation is
expected to be consistent. The coincidence, however, might be 
only fortunate. A complete calculation that would crossover
from the $\kappa^{-2}$ behaviour of the electrostatic stiffness
which is well understood for rigid chains, to the $\kappa^{-1}$ behaviour
that is suspected for flexible chains on the basis of variational
calculations, is still missing.

{\it 4.3 The case of poor solvents}

The above discussion focuses on the case where the neutral chain backbone
is in a good or theta solvent. A rather different result is obtained for 
the case of a flexible, weakly charged chain in a poor solvent. This case was
considered in the absence of screening in section (2.1), where the chain
 was described as a linear string of "poor solvent blobs", each of size
$\xi_e=b(f^2\lb/b)^{-1/3}$ and containing $ \tau f (f^2\lb/b)^{-1}$ charged monomers
(cf. equation (\call{cblob2}). Following the procedure of \refto{KK82},
one can attempt to apply the OSF theory to the chain of blobs. The theory 
is valid if the linear charge density along the chain of blobs,
$fg/\xi_e$, is larger than the critical value $(\xi_e\lb)^{-1/2}$. 
This condition can be rewritten as $\tau > (f^2\lb/b)^{1/3}$. For weakly
charged chains ($f<<1$), it is thus satisfied even for moderately poor 
solvents.
The electrostatic persistence length that results from applying the OSF theory
is
$$\eqalign{
\ell_p &= \xi_e +{ f^2 g^2\lb \over \kappa^2 \xi_e^2 }\cr
       &= \xi_e + ({\tau^2 \over b \kappa^2}) ({f^2 \lb \over b })^{-1/3} \ .
}\eqno(lpoor)
$$
The surprising implication of this result is that the persistence length,
(and therefore the radius of gyration of the chain), increases as the 
solvent quality decreases. This increase is the consequence of an increase
in the linear charge density, which is itself induced by the collapse 
of the monomers into dense blobs. This effect, however, exists 
only as long as the charge density exceeds the value $\lb^{-1}$ 
at which Manning condensation takes place. This gives an upper limit for
$\tau$, $\tau < f (f^2\lb/b)^{-1/3}$. The conclusion is that an increase
in the persistence length as $\tau$ increases  should occur
in the temperature range 
$ f^{1/3} (b/\lb)^{1/3} > \tau > f^{2/3} (\lb/b)^{2/3} $. 

\bigskip

{\bf 5. Charged Gels and Brushes}

\taghead{5.}

\smallskip

Many properties of polyelectrolyte solutions are dominated not by 
the chain conformation, but by the counterions. We discuss below
 two examples of this situation,
the polyelectrolyte gel 
and the polyelectrolyte brush. In both problems, the polyelectrolyte chains 
occupy a small region of space (close to the grafting surface for the brush), 
surrounded by pure solvent. The small ions in the solution are free 
to diffuse in and out the polyelectrolyte region,
 and the external solvent region acts as 
a reservoir that imposes the small ion  chemical potential. 
A Donnan equilibrium \refto{Hill,Donnan} is reached,  that relates 
the concentrations of the small ions inside the polyelectrolyte region to their 
imposed  concentration in the reservoir. If the monomer concentration 
in the polyelectrolyte region is $c$, the charge density due to the polymer is $fc$. The reservoir 
contains salt at a density  $n$, and the Debye-H\"uckel screening length $\kappa^{-1}$ is given by 
$\kappa^2= 8\pi n \lb$. For simplicity we assume in the following that the counterion of the polymer 
is identical to the positive ion of the salt, we call $n_+$ and $n_-$ the concentration of positive  
and negative ions respectively in the polyelectrolyte region. 

The polyelectrolyte charge creates an electrostatic potential 
difference between the polyelectrolyte region and the reservoir
$U$. The chemical potentials of the small ions in the polyelectrolyte region 
are $\mu_+=k_B T  \ln n_+ +U$ and   
$\mu_- =k_B T  \ln n_- -U$. At equilibrium these chemical potentials are equal 
to the salt chemical potential in the reservoir  $\mu =k_B T  \ln n$. 
This leads to a chemical equilibrium law for the small ions
$$\eqalign{
n_+ n_- =n^2
}\eqno(donnan)
$$
A second relation between the concentrations $n_+$ and $n_-$ is provided by the electroneutrality 
condition $n_+ =n_- +fc$. The salt concentration inside the polyelectrolyte region 
(given by $n_-$) is smaller than that of the reservoir. The difference 
in osmotic pressure between the two regions is 
$\pi = k_B T ( n_+ + n_- -2n) = [(fc)^2 + 4n^2]^{1/2} -2n$.
When the salt concentration $n$ is smaller 
than the counterion concentration $fc$, the difference in salt concentration 
is small and the osmotic pressure is that of the ideal gas 
of counterions discussed in section (3.1),
$\pi = k_B T fc$. In the limit where the salt concentration is larger 
than the counterion concentration,  the osmotic pressure is given by 
$$\eqalign{
\pi = k_B T {(fc)^2 \over {4n}}= k_B T {c^2 \over 2} {{4\pi\lb f^2}\over {\kappa^2}}
}\eqno(donnan2)
$$
This osmotic pressure defines an effective virial coefficient between monomers 
$v_{el}={f^2 / {2n}}= {{4\pi\lb f^2} {\kappa^{-2}}}$. 
This is identical to
the excluded volume that can be calculated from the Debye-H\"uckel interaction
(\call{vdh})
between monomers.

The major assumption made here is that the ion densities are uniform. 
This is the case when regions of size $\kappa^{-1}$ 
around each polyelectrolyte,  where the counterion concentration is increased, overlap,  i.e. 
when the Debye-H\"uckel screening length is larger than the distance between chains. 
This is always true in the absence of salt. The osmotic pressure calculated from the 
Donnan equilibrium is of purely entropic origin.
 If the ion densities are uniform the electrostatic 
contribution to the osmotic pressure is small compared 
to the translational entropy of the small ions
as checked below for a gel. The osmotic pressure is thus independent 
of the strength of the electrostatic interaction 
characterised by the Bjerrum length $\lb$. 
The only role of the electrostatic interaction is to enforce
 electrical neutrality.

{\it 5.1 Grafted polyelectrolyte layers} 

Grafted polymer layers (polymer brushes)  \refto{Milner}
are obtained by attaching one 
of the polymer end points on a planar solid surface. If the distance between 
chains is smaller than their natural size,  a thick layer with a size 
proportional to the chain molecular weight is formed. The chains are stretched by 
the repulsive interactions between monomers. 

Polymer brushes are often very efficient 
to enhance colloidal stabilisation \refto{Napper}.
 Grafted polyelectrolyte layers 
have been described theoretically by several authors 
\refto{PI91,ZB92,Miklavic,Indiens}
we follow here the lines of the original work of Pincus
which assumes that the polymer concentration is uniform throughout 
the thickness $h$ of the brush. A similar approximation for neutral 
polymer brushes was introduced by Alexander \refto{AL79} and
 de Gennes \refto{dgbrush}.
 We also assume in this section that 
the local electrostatic screening length is larger than the distance 
between chains imposed by the 
distance $D$ between grafting points so that the counterion
 concentration is roughly constant in the layer.

The electrostatic field and the distribution of counterions
in the vicinity of a charged solid surface with a surface charge density 
$\rho$ can be determined from the
Poisson-Boltzmann equation. In the absence of added salt, 
the counterion density decays as a power law of the distance 
from the surface, and the counterions are confined in the vicinity of the surface
over the Gouy-Chapman length $\lambda=(2\pi \rho \lb)^{-1}$. If the grafting surface 
is neutral, the charge density of a grafted polyelectrolyte with 
$\sigma=D^{-2}$ grafted chains per unit area
layer is due to the monomer charges $\rho= \sigma Nf$, the Gouy-Chapman length 
\refto{IS85}
of the polymer brush is 
$$\eqalign{
\lambda \simeq {1\over { \sigma f N \lb}}
}\eqno(gc)
$$
Two cases must then be considered. When the Gouy-Chapman length is larger than the brush size $h$,
most of the counterions are outside the brush, and the brush is charged. 
When $\lambda$ is smaller than the brush height, the counterions are confined in the grafted layer
and the brush is neutral.

In the neutral limit, the brush is swollen by 
the osmotic pressure of the counterions $\pi = k_B T fc$. 
This pressure is balanced by the elasticity of the polymer chains. For a gaussian chain, 
the elastic pressure is $\pi_{el}={{\sigma h}/ (Nb^2)}$. The monomer 
concentration in the brush being $c=\sigma N/h$ 
the thickness of the brush is given by 
$$\eqalign{
h=N b f^{1/2}
}\eqno(nb)
$$
The thickness of the brush is in this regime independent of the strength 
of the electrostatic intractions since the stretching of the chains 
is due only to entropic effects. The thickness given by equation  (\call{nb}) 
is larger then the size of an individual chain in a dilute solution  given by (\call{R1}). 
This result has been obtained by assuming that the monomer
 concentration is constant
inside the adsorbed polymer layer. 
Similar scaling results are obtained if this constaint is released. 
Using a self-consistent field method, Zhulina et al. \refto{ZB92} have shown 
that the concentration profile has then a Gaussian decay.

In the charged limit, one can in a first approximation assume that 
the counterion concentration is constant up to the Gouy-Chapman length,
 it is equal to
$c_i=  (f c h )/ \lambda$. The counterion 
pressure that stretches the chain is therefore
$\pi= k_B T fch /  \lambda$. The thickness of the brush is then
$$\eqalign{
h=N^3 \lb b^2 \sigma  f^2
}\eqno(cb)
$$
In this regime the electrostatic interactions have a contribution of 
the same order as the entropic contribution and the thickness 
depends on their strength. The thickness of the layer grows 
faster than linearly with molecular weight, however one should
 keep in mind that 
the layer thickness remains smaller than the Gouy-Chapman length and thus that this 
thickness remains smaller than the thickness in the neutral regime (\call{nb}). 
A precise determination of the counterion profile in this regime has been done by Pincus \refto{PI91}, 
the same scaling laws are obtained when this profile is taken into account.

When salt is added to the layer, the polyelectrolyte brush behaves as a neutral polymer brush  
with an effective excluded volume $v_{el}={f^2 / (2n)}$;  the thickness of the brush is then
$$\eqalign{
h=Nb  (\sigma b^2)^{1/3}  f^{2/3} (nb^3)^{-1/3}
}\eqno(bs)
$$
The monomer concentration profile has also been calculated in this limit by Zhulina et al. \refto{ZB92}
The important result is that the cross-over to the neutral brush regime (\call{nb}) 
occurs when the salt concentration is larger than the counterion concentration in the brush. 
This leads in general to a high ionic strength and the thickness 
of a grafted polymer layer is rather insensitive to ionic strength 
over a broad range of ionic strength. This makes grafted polyelectrolyte 
layers particularly interesting to promote colloidal stabilisation.

If the polyelectrolyte is annealed, the pH 
inside the grafted layer can be significantly different from outside.
This could induce a non monotonic variation of the thickness with 
grafting density \refto{Borisov}.

{\it 5.2 Polyelectrolyte gels}

A model based on the Donnan equilibrium, very similar to that of polymer 
brushes, can be made for charged polymeric gels \refto{BJ92}.
 The polymer gel is 
at equilibrium with a reservoir of solvent if the osmotic pressure in the gel is equal
to the osmotic pressure of the salt in the reservoir, i.e. 
if the swelling osmotic pressure due to the entropy of the counterions is balanced by the 
pressure due to the elasticity of the polymer chains. 
If the chains between the crosslinks of the gel are gaussian chains and have N monomers 
and an end to end distance (mesh size) $R$, the elastic pressure is of order
 $\pi\simeq k_B T  (c/N) (R^2 / Nb^2)$. If we now make the $c^*$ 
assumption proposed by de Gennes \refto{GE79} that,
in the swollen gel at equilibrium, the chains are just at the overlap concentration, 
the mesh size of the gel is such that $c=N/R^3$. In the absence of salt, 
the mesh size is then given by equation (\call{nb})
$$\eqalign{
R=N b f^{1/2}
}\eqno(gm)
$$
In the presence of salt, the mesh size is given by
$$\eqalign{
R=N^{3/5} b ({f^2 \over {2n}})^{1/5}
}\eqno(gs)
$$
The important point to notice is that, in contrast to neutral gels, 
the mesh size of the gel is different 
from the radius of an isolated polymer chain in a dilute solution. As explained above,
these results are valid when the salt concentration 
is small enough that the Debye-H\"uckel screening length is larger than the gel mesh size i.e. 
when the mesh size is larger than the radius of an isolated chain in solution given by (\call{R}). 
The chains in this case should be viewed as chains under strong tension. 
The small ions osmotic pressure is exerted at the surface of the gel and  the force is transmitted 
to the internal chains via the crosslinks.
 
The mesh size and thus the equilibrium swelling 
of the gel do not depend on the strength of the electrostatic 
interaction $\lb$. When the monomer 
concentration in the gel is uniform, the electrostatic energy of the gel vanishes. 
The actual concentration in the gel is not homogeneous and 
the inhomogeneity can be characterised by 
the structure factor $S(q)$. The electrostatic 
energy per unit volume of the gel can then be expressed as 
$$\eqalign{
E_{el}= c {f^2\over 2} \int d\qq S(q){{4\pi \lb}\over {q^2+\kappa^2}}
}\eqno(gelen)
$$
The structure factor of the charged gel is not known exactly but one can 
expect it to have a sharp peak at a position $q*\simeq  R^{-1}$ 
the height of the peak being of order $N$. The integral giving the electrostatic energy
of the gel is dominated by the peak of the structure factor 
and can be estimated up to numerical prefactors
$$\eqalign{
E_{el}= {ck_B T \over N} {{f^2 N^2 \lb}\over R}{1\over {1+{\kappa^2R^2}}}
}\eqno(gelen')
$$
It can then be direcly checked that, in the weak screening limit where $\kappa R<<1$, 
the electrostatic energy is smaller than the translational entropy of the counterions.

In the strong screening limit $\kappa R>>1$, the electrostatic interactions are dominant.
The charged monomers in the gel interact via a Debye-H\"uckel potential 
and the electrostatic interactions 
contribute to the rigidity of the chains between 
crosslinks as discussed in the previous section. 
At very high ionic strength the electrostatic persistence length 
is small, and the chains behave 
as neutral chains with an electrostatic excluded 
volume $v_{el}={f^2 \over {2n}}$. 
The same scaling result as equation (\call{gs})
 is then found for the mesh size.

The elastic shear modulus of the gel can be calculated by imposing a 
uniform deformation $\gamma$ to the crosslinks. The increase of the square of the mesh 
size imposed by the deformation is of order $(\gamma R)^2$ and 
if the chains are gaussian, the elastic energy stored in the gel is
$$\eqalign{
F_{el}=k_B T  {c\over N} {\gamma^2R^2 \over {Nb^2}}
}\eqno(elast)
$$
This leads to a shear modulus 
$$\eqalign{
G=k_B T  {c\over N} {R^2 \over {Nb^2}} & \simeq k_B T  fc \ \ \ ({\rm salt\ free\ case})\cr
 &\simeq k_B T  {{f^2c^2} \over n}\ \ ({\rm  with\ added\ salt})
}\eqno(elast)
$$
In all cases, the shear modulus of the gel is thus of the order of the osmotic 
pressure of the small ions (counterions and salt)

The last property of the gel that we discuss is the cooperative 
diffusion constant, that can be obtained from a two-fluid model
\refto{tanakabenedek}. The displacement field of the gel
$u(\rr,t)$ results from a balance between the elastic restoring 
force and a viscous force due to the solvent drag. The elastic force per 
unit volume of the gel is $G\nabla^2 u$. 
The viscous force exerted by a solvent flowing through a porous medium of pore size $R$ is 
given by the Brinkman equation  \refto{BR47} 
and is of order $\eta R^{-2} {{\partial u}\over {\partial t}}$ 
where $\eta$ is the solvent viscosity. 
The force balance on the gel is then written as 
$$\eqalign{
\eta R^{-2} {{\partial u}\over {\partial t}}= G\nabla^2 u
}\eqno(fb)
$$
The relaxation of the gel deformation is diffusive with a cooperative diffusion constant
$$\eqalign{
D= {{GR^2}\over \eta}
}\eqno(dcoop)
$$
The cooperative diffusion constant is independent of the 
molecular weight in the absence of salt ($D\simeq f^{1/2}$). 
It decreases with $N$ in the presence of salt ($D\simeq N^{-2/5} (f^2/n)^{1/5}$).

Experimentally, the properties of charged gels, as well 
as the properties of neutral gels,
 strongly depend on the preparation condition of the gels. The simple results presented here 
(based on the so-called $c^*$ theorem) implicitly assume 
that the gel was prepared by crosslinking chains in the
 vicinity of their overlap concentration. 
A more refined theory \refto{RubinsteinJoanny} based on an affine deformation hypothesis first
 proposed by Flory \refto{Flory} and then developped
 by Panyukov \refto{Panyukov} has recently been constructed. It provides an 
explanation for the experiments of Candau and coworkers \refto{Candau}
 which show that the elastic shear modulus of 
polyelectrolyte gels with a fixed concentration smaller than 
the equilibrium swelling concentration decreases 
with the fraction of charged monomers.

\smallskip

{\bf 6. Semidilute solutions}

\taghead{6.}

The overlap concentration of polyelectrolyte solutions in the absence of salt is very low; 
the chain radius given by equation (\call{R}) increases linearly with molecular weight and 
the overlap concentration decreases as $c^*\simeq b^4/(\lb N^2f^2)$. 
Most of the experiments with long chains
are thus performed in the semidilute range where the electrostatic 
interactions between different chains are strong. Our understanding of the conformation 
of polyelectrolyte chains in semidilute solutions is however rather 
poor and no general view is available. We discuss in this section a few aspects of the static 
properties of interacting polyelectrolyte chains.

{\it 6.1 Ordering transitions in polyelectrolyte solutions}

The models presented in section 2 to describe 
the conformation of polyelectrolytes in salt free dilute solutions consider 
only a single chain and ignore the screening due to the counterions. If the polymer
concentration is finite, the counterion concentration is finite and the screening 
length due to the counterions $\kappa^{-1}$ is given by $\kappa^{2}=4\pi \lb fc$. 
In a dilute solution the screening length is always larger 
than the average distance between chains $d \simeq (N/c)^{1/3}$.
Different chains thus interact via a long range pure Coulomb potential. The polyelectrolyte 
chains are strongly charged objects and this long range interaction can lead to the 
formation of Wigner crystals, very similar to the colloidal crystals observed 
in solutions of charged spherical particles \refto{SafranChaikin}
or charged elongated particles (viruses) 
with a mesoscopic size. 
The centers of mass of the particles are then regularly distributed on a periodic lattice. 
The solution 
has a tendency to crystallise when the interaction 
between neighboring chains $k_B T  N^2f^2 \lb /d$ is 
larger than the thermal excitation $k_B T $. This naive estimate 
predicts crystal formation at a very 
low concentration, way in the dilute range $c \simeq 1/(N^5f^6\lb^3)$. The crystal is expected 
to melt when screening is important, in the vicinity of the overlap concentration.
There is no clear experimental evidence for this crystallisation 
in flexible polyelectrolyte solutions. 
This may be due to the weak elastic resistance of this crystal that could be 
destroyed by any kind of perturbation (mechanical perturbation). 
In any case we expect strong interactions in dilute polyelectrolyte 
solutions that lead to a structuration of the solution and thus to 
the appearance of a peak in the structure factor $S(q)$ of 
the solution corresponding either to liquidlike or to solidlike order. The wavevector $q^*$ at
 the peak is of the order of the inverse distance between chains 
$q^* \simeq 2\pi /d \simeq (c/N)^{1/3}$ \refto{Kajijphys}. The peak is observed 
experimentally and the variation of its position with concentration
is in good agreement with this prediction. It can be suppressed by adding
salt to the solution, so that the screening length $\kappa^{-1}$
becomes smaller than the distance between chains.

As in dilute solutions, polyelectrolytes have, in a semidilute solution, 
a locally rodlike conformation that can be 
characterized by a persistence length $\ell_p$.
Quite similarly to rodlike molecules, semiflexible 
macromolecules undergo an Onsager 
transition \refto{liqcryst,Onsager} between an isotropic liquid phase and 
a nematic ordered phase where the molecules are parallel. 
The Onsager concentration where this transition takes place 
has been calculated by Semenov and Khokhlov \refto{semenov,grosbergkhokhlov},
 for neutral molecules of diameter $d$ it scales as
$c_o\simeq 1/(\ell_p db)$. For charged molecules, the diameter of the molecule must be replaced by 
an effective diameter equal to the range of the electrostatic interaction $\kappa^{-1}$ \refto{Onsager} and the 
Onsager concentration is of order $c_o\simeq 1/(\ell_p \kappa^{-1} b)$. All the models for 
polyelectrolyte conformation based on the Odijk-Skolnick-Fixman approach 
predict an isotropic-nematic transition for polyelectrolyte solutions in the 
presence of salt. Experimentally, the transition is
only observed for very rigid polymers (such as DNA where a cholesteric phase is observed) 
where the intrinsic persistence length is very large and dominates the 
electrostatic persistence length. There is no definite evidence for an Onsager 
transition in flexible polyelectrolyte solutions. This is an unresolved issue.
 Possible explanations are based on the idea that the scaling theories 
ignore some prefactors that may be large and shift the 
transition to an unobservable value. The isotropic-nematic transition is also expected only for 
anisotropic enough objects, the ratio between the persistence length 
and effective diameter $\kappa^{-1}$ must be larger than 
a finite number $ \simeq 5$ \refto{frenkel}. This criterion is not 
always met experimentally. A thorough theoretical study of possible isotropic-nematic 
transition in polyelectrolyte solutions has been made by Nyrkova \refto{nyrkova}. In the following 
we will ignore this possibility and consider that a semidilute polyelectrolyte solution 
remains isotropic at any concentration.

\smallskip

{\it 6.2 Correlation length and osmotic pressure of semi-dilute polyelectrolyte solutions}

In a dilute solution, polyelectrolyte chains have a rodlike conformation 
at a local scale smaller than the persistence length $\ell_p$ and the 
persistence length is always larger (or of the same  order of magnitude)
than the screening 
length of the electrostatic interactions. In a semidilute solution, 
the polymer chains overlap and, if we assume that they do not form ordered phases, 
they form a temporary network with a mesh size $\xi$. In the absence of salt, the chains also 
have a rodlike conformation at the scale $\xi$ and 
their persistence length is, as shown below, 
larger than the mesh size. The mesh size can then be calculated from a scaling argument by imposing
that it is equal to the isolated chain radius (\call{R}) at the overlap concentration $c^*$. 
(For simplicity,
 we suppose here that locally the polymer shows local gaussian statistics, 
the other cases can be treated in a similar way). Equivalently, a pure geometrical argument
can be used, by imposing a close packing condition for chain subunits of size $\xi$, 
$c=(\xi/\xi_e) g / \xi^3$. In any case, $\xi$ is given by \refto{GP76,PF78}
$$\eqalign{
\xi \simeq b^{-1} c^{-1/2} \xi_e^{1/2} 
}\eqno(xi)
$$
The number of monomers within a volume   of size $\xi^3$, or correlation blob, 
 is then 
$G_\xi= c^{-1/2}\xi_e^{3/2} b^{-3}$.

For weakly charged polyelectrolytes, the counterions are free and
 behave roughly as 
an ideal gas. The screening length due to the counterions is then  $\kappa_i^{-1}=(4\pi \lb fc)^{-1/2}$. 
This screening length is of the order of the mesh size for strongly charged polyelectrolytes
 $(f \simeq 1)$ but it is larger than $\xi$ for weakly 
charged polyelectrolytes. The polymer itself, however, 
 contributes to the screening of the electrostatic 
interactions and the actual screening 
length is smaller than the counterion screening length. From a purely electrostatic
 point of view each cell of size $\xi $ 
containing $G_\xi$ monomers is neutral and can be 
considered as independent of all the others.
 It is thus reasonable to assume that the effective 
screening length is of the order of the mesh size $\xi$ \refto{KK82,DC95}.
\note{This is somewhat similar to the assumption that, in neutral polymers,
the {\it hydrodynamic} screening length is equal to the static
correlation length.}

As salt is added to the solution, the 
screening length decreases and becomes dominated by salt 
($\kappa^{2}= 8\pi n \lb$) when the Debye-H\"uckel screening length of 
the salt $\kappa^{-1}$ is smaller than 
the mesh size $\xi$. When the screening is dominated by the salt, the electrostatic persistence 
length of the chains decreases with the salt concentration. 
As long as  the persistence length is larger
than the mesh size $\xi$, 
the mesh size is still given by equation (\call{xi}). 
If the persistence length is smaller than 
the mesh size, the polyelectrolyte solution 
behaves as a neutral semi-flexible polymer 
solution  and the mesh size or correlation length 
can be calculated from the radius in a dilute solution using scaling arguments. 
If the Odijk-Skolnick-Fixman statistics is assumed \refto{OD79,KK82}
$$\eqalign{
\xi \simeq b \xi_e c^{-3/4} \kappa^{3/4} b^{-3/2},
}\eqno(xikk)
$$
while if  the persistence length is assumed equal to the screening length
\refto{GP76,PF78}
$$\eqalign{
\xi \simeq\xi_e^{3/4} c^{-3/4} \kappa^{1/2} b^{-3/2}
}\eqno(xiP)
$$

Experimentally, the important feature of semidilute 
polyelectrolyte solutions is the fact that, if the salt concentration is low enough, 
the structure factor $S(q)$ has a  peak at a finite wavevector $q^*$. At small wavevectors, 
the structure factor, because of the electroneutrality of the solution, 
is dominated by the small ions,  that behave as an ideal gas. 
In the absence of salt, $S(q=0)=1/f$. If the salt concentration
 n is larger than the counterion concentration, $S(q=0)=2n/(f^2c)$. 
At large wavevectors, corresponding 
to distances smaller than the mesh size $\xi$, the polyelectrolyte 
has a rodlike behavior and the structure factor is given by 
$S(q)=\xi_e /(qb^2)$. When 
the wavelength $q^{-1}$ is equal to the mesh size, the value of the structure factor is 
$S^*=G_\xi= c^{-1/2}\xi_e^{3/2} b^{-3}$. It can be checked explicitly 
that this value is larger than the thermodynamic value 
$S(q=0)$ and the  structure factor must thus have a peak at a finite
 wavevector $q^*$. The existence of the peak is therefore related to the very small 
compressibility of the small ion gas which gives $S(q=0)$
due to 
the electroneutrality constraint. 
This discussion is valid as long as the counterions and 
the salt ions are uniformly distributed throughout the solution,
 i.e. as long as the screening length 
 is larger than the mesh size of the solution. In the salt dominated regime,
 where the salt 
screening length is smaller than the mesh size of the solution, 
the counterions are 
confined in a sheath of size $\kappa^{-1}$ around each polymer 
and the solution essentially behaves as a neutral polymer solution 
where the structure factor decays monotonically. 
The peak thus disappears in the salt dominated regime. At the crossover between the two regimes, 
the $q=0$ value of the structure factor increases very sharply. 
When the effect of salt is not dominant, the peak 
position $q^*$ defines the correlation length 
of the solution and one expects that it is 
of the order of the inverse of the mesh size $q^*\simeq 1/\xi$. The position 
of the peak of the structure factor increases thus as $c^{1/2}$ in a semidilute solution (and as
$c^{1/3}$ in a dilute solution as explained above). 
It is also important to note 
that the correlation length $\xi$ and thus the peak position are roughly 
independent of the salt concentration as long as the Debye-H\"uckel 
screening length is smaller than $\xi$. These results are 
in rather good agreement with experiments. 

In the osmotic regime where the Debye-H\"uckel screening length is smaller 
than the mesh size, the small ions are uniformly distributed and
the osmotic pressure of the solution is the same as that of a gel as discussed in section 5, it
is dominated by the counterions. It is equal to $\pi =k_B T fc$ if the counterion 
concentration is smaller than the salt concentration, 
($fc>n$) and to $\pi =k_B T f^2 c^2 /(4n)$ if $fc<n$. 
In the salt dominated regime $\kappa^{-1}<\xi$, 
the polyelectrolyte 
chains behave as neutral semiflexible chains with a persistence length $\ell_p$ and 
an effective diameter equal to the screening length $\kappa^{-1}$ (up to logarithmic corrections).
The excluded volume between Kuhn segments of size $\ell_p$ is of order
 $\ell_p^2 \kappa^{-1}$. If the persistence length is larger than the mesh size
(or of the same order), 
the osmotic pressure varies as
$$\eqalign{
\pi/ k_B T= c^2 \kappa^{-1} \xi_e^{-2} b^4
}\eqno(osmP)
$$
At the crossover between the osmotic and the salt dominated 
regime, the osmotic pressure varies 
very rapidly and the crossover is not smooth. 
When the persistence length becomes small enough compared
 to the mesh size, the excluded volume 
correlations are relevant and the osmotic pressure increases 
as $\pi \simeq c^{9/4}$.
At very high ionic strength, the electrostatic 
interaction is very weak and cannot be approximated by an excluded volume between Kuhn segments. 
In this weak coupling regime
 there no longer is an electrostatic persistence length 
(it is shorter than the screening length) and the polyelectrolyte 
chains should be considered as flexible chains
 where the monomers interact via the short range
 Debye-H\"uckel potential with an effective excluded volume $v_{el}= f^2/2n$. 
The smooth crossover between the salt dominated and 
the weak coupling regimes occur when 
$\kappa \xi_e \simeq 1$ i.e. when the screening length 
becomes equal to the electrostatic blob size.

{\it 6.3 Electrostatic rigidity in semidilute solutions}

In this section we discuss the conformation of polyelectrolyte chains 
in a semidilute solution and the effect of the interactions between different chains on the 
persistence length of a polyelectrolyte. For simplicity we consider only a 
strongly charged polyelectrolyte for which the bare persistence length $\l0$  
and the distance between charges $A$ are such that $A^2>\l0\lb$ so that the Odijk-Skolnick-Fixman 
theory can be applied in a dilute solution.
 We also assume that the bare persistence length is much smaller 
than the mesh size of the semidilute solution. The results can easily be extended 
to weakly charged polyelectrolytes if necessary by renormalising 
the bare persistence length to the blob size.

Two types of theories have been proposed to describe chain conformation in a semidilute solution.
The early work of Odijk \refto{OD79,Brochardetal} makes the assumption that 
the interchain interactions have a negligible influence on 
the persistence length so that the persistence length is given by the usual
Odijk-Skolnick-Fixman theory with the relevant screening length. A scaling theory for 
the chain conformation has been built on this assumption. It was however later
pointed out by Witten and Pincus \refto{WP87} that if a polyelectrolyte solution is isotropic, 
the electrostatic interaction energy between two chains can be reduced by a bending of the chains 
to avoid each other and thus that the interaction energy between 
different chains reduces the persistence length. A persistence length equal 
to the mesh size of the solution is obtained in the osmotic regime where the Debye-H\"uckel 
screening length of the salt is larger than the solution mesh size. 
In a revised version of this theory, in the salt dominated regime, 
the corrections to the persistence length due to the interactions
 between chains are found to be small.

The simplest approach \refto{BJ94} to the persistence length of a polymer 
chain in a semidilute solution is to use linear response theory. 
Within the framework of linear response, an effective pair interaction 
between monomers can be introduced. The Fourier transform of the effective interaction is
related to the structure factor S(q) of the solution by 
$\hat{v}_{eff}(q)= \hat{v}_{DH}(q)
 \left(1-cS(q)\hat{v}_{DH}(q)\right)$. Following the lines of Odijk argument, 
the effective persistence length can then be calculated as
$$\eqalign{
\ell_p = \l0 + { \hat{v}_{DH}(q=0) \over 16 \pi k_B T A^2} 
\left(1-c S(q=0) \hat{v}_{DH}(q=0)\right).
}\eqno(lplin)
$$
This relation gives the persistence length as a function of the $q=0$ value of 
the structure factor which is a thermodynamic quantity. If the explicit 
form of the Debye-H\"uckel potential is substituted, the persistence length is 
$\ell_p = \l0 + \lo(1-4\pi\lb cS(q=0)/\kappa^2 )$. A reduction of the persistence length 
from the single chain value induced by interchain interactions is therefore predicted.
However this result can be used only in the limit of linear response, 
i.e. if the interaction between chains is smaller than $k_B T $. The interaction between two rods 
crossing at perpendicular angle is 
$$\eqalign{
\beta = {2\pi\lb \over \kappa A^2};
}\eqno(defb)
$$
The linear response result is thus valid if $\beta<<1$ which corresponds 
to the weak coupling regime for rigid polyelectrolytes.

No direct calculation of the  effect of interchains interactions on
the persistence length seems available 
in the strong coupling regime $\beta>>1$. 
Some insight into the problem can however be gained by looking at the following 
simpler two dimensional problem:  one 
semiflexible test chain lies in a plane, and interacts with all the other 
chains replaced by infinite rods perpendicular 
to this plane with a concentration per unit area $\Gamma$. 
It is a good approximation to replace the other chains by infinite rods
as 
long as their persistence length is larger than the screening length. 
In the limit where the Debye screening length is smaller than the distance between the chains, the
solution can be considered as a quasi ideal gas of rods. When the test chain bends, 
the distribution of the rods around the test chain is changed and this contributes to 
the bending energy. 
The persistence length can be calculated following the lines of the work of 
Odijk in a dilute solution,
the result is  
$$\eqalign{
\ell_p = \lo -{\Gamma \kappa ^{-3} f(\beta)}
}\eqno(lpob)
$$
where the function $f(\beta)$ is defined as 
$$\eqalign{
f(\beta)= {\beta \over 2}
 \int_0^{+\infty} {\rm d}z \exp \left(-\beta \exp\left(-z\right)
 \right) \left[
-\beta z^2 \exp(-2 z) + \exp(- z) (-z^2 + z
+ 1 )\right]
}\eqno(fbeta)
$$
The correction to the Odijk value is always negative 
 which corresponds also to a reduction of the persistence length from the isolated chain value.
In the weak coupling limit,  $f(\beta)=3\beta^2/8$ in agreement with the linear response theory. 
In the strong coupling limit, the interaction is very similar to a hard sphere interaction and 
$f(\beta)= (\ln \beta)^2$ varies only logarithmically with the interaction strength. 
The correction to the single chain persistence length due to interchain interactions is small.

These results obtained in two dimensions find a simple interpretation in terms of individual
deflections of the test chain by the perpendicular rod. When the 
test chain interacts with one rod at a distance $r$, 
it bends and its orientation changes over a distance $\kappa^{-1}$ by a finite angle
that can be shown to be equal to
$$\eqalign{
\theta \simeq \kappa r \exp (-\kappa r).
}\eqno (defa)
$$
The persistence length can then be obtained by summing the individual deflections 
weighed by the Boltzmann factor related to the direct electrostatic 
interactions between the chains. 
The persistence length given by equation (\call{lpob}) is then found up to a numerical prefactor. 
The independent deflection model is easily generalisable to real three dimensional chains. 
The chains do not always cross at a finite angle,
 but the conformations 
where two chains are almost parallel are energetically costly, 
 since the electrostatic interaction 
is very high in this case (it increases linearly with chain length). 
This leads to a persistence
length
$$\eqalign{
\ell_p^{-1} = \lo^{-1} \left[1 + \ln^2(\beta) \left(\alpha c A^3 \over \kappa \lb\right)
\right].
}\eqno(ldef3d)
$$ 
where $\alpha$ is a numerical constant of order unity. The important result 
is that the correction is small and thus that in the salt dominated regime, 
the reduction of the persistence length due to 
the interactions between different chains can be neglected. 

The situation is very different in the osmotic regime, the chains 
interact very strongly and the distance between interacting chains is always of the order of the
mesh size $\xi \simeq (cb)^{-1/2}$; the screening length $\kappa^{-1}$
is also of order $\xi$. There is no precise theory of the persistence length in the osmotic regime, 
the two following qualitative arguments suggest however that the persistence 
length is of the order of the mesh size $\xi$. 
As the distance $r$ between interacting chains is of order $\kappa^{-1}$, 
the deflection angle given by equation (\call{defa}) is always of order one.
Whenever two chains cross they bend by an angle of order 
unity and loose the memory of their orientation. The persistence length is thus of 
the order of the distance between crossings i.e. of the order 
of the mesh size $\xi$. The second argument 
is based on the strong structuration of the solution in the osmotic regime. 
The structure factor has a strong peak and with a very rough approximation the chains can be thought 
of as lying on a lattice with a lattice constant of order $\xi$. The total electrostatic energy of 
this lattice does not depend of the bending of 
the chains over distances larger than $\xi$. When the chains are bent on a scale 
$\xi$, entropy is gained but at the expense of bending energy 
associated with the bare persistence length $\l0$. The entropy is clearly dominant when $\l0<\xi$ 
which leads to a persistence length of order $\xi$.
An important consequence is that in saltfree semidilute
solutions, both flexible 
and rigid polyelectrolytes have a persistence length of the order
of the correlation length.
 It is also important to notice 
that the persistence length of rigid polyelectrolytes is predicted to
vary  non-monotonically with ionic strength. 
In the osmotic regime, it is dominated 
by interchain interactions and increases with ionic strength. 
In the salt-dominated regime, it is dominated 
by intrachain interactions and decreases with ionic strength. There is no clearcut 
experimental evidence of this non monotonic variation of the persistence length with ionic strength \refto{OostwalOdijk}. 

{\it 6.4 Concentrated  solutions of flexible polyelectrolytes}

The semidilute regime discussed in the two previous subsections 
corresponds to chains which are rodlike at intermediate length scales and 
that can be described locally by the electrostatic blob model of section 2. At a
higher concentration however, the mesh size of the solution is smaller 
than the electrostatic blob size. Electrostatic interactions 
only play a minor role in this regime and the polyelectrolyte 
chains remain gaussian at all length scales. The electrostatic blob size $\xi_e$ 
is equal to the correlation length $\xi$ when the electrostatic blobs are close 
packed. If the local structure of the blobs is gaussian, this occurs at a concentration
$$\eqalign{
c^{**}=b^{-3} f^{2/3} (\lb /b)^{1/3}
}\eqno(conc)
$$
In the concentrated regime $c>c^{**}$, the elecrostatic 
interactions are weak and the correlations in the solution can be
 studied within 
the framework of the so-called random phase approximation, as first 
suggested by Borue and Erukhimovich \refto{BE,JL91}. 
This leads to a structure factor 
(monomer-monomer concentration correlation function)
$$\eqalign{
{1\over S(q) }
= {1 \over S_0(q)} + v + w^2 c + 
{4 \pi \lb f^2 \over q^2 +\kappa^2} \ .
}\eqno(rpa)
$$
Here $S_0(q)$ is the structure  factor of gaussian chains,
given by the Debye function, and in the following
it is written 
 as $S_0(q)^{-1}=(Nc)^{-1} [1+ g(Nq^2b^2/6)]$.
The second and third virial coefficients
that describe the interactions between {\it neutral} monomers (i.e. that would
describe the
interactions for $f=0$) are, respectively, $v$ and $w$. 
 The second virial coefficient $v$  is the excluded 
volume $v$ and vanishes at the $\theta$ temperature. 
The third virial coefficient $w^2$ 
is assumed here to be positive.
The last term comes from the screened electrostatic interactions. The 
screening is due to all the small ions 
and the Debye-H\"uckel screening length is defined here  
by $\kappa^2= 4 \pi \lb(fc+2n)$.   The statistics 
of the electrostatic blobs remain 
gaussian if the excluded volume is small enough,  namely if 
the electrostatic blob size $\xi_e$ is smaller than the so-called 
thermal blob size $\xi_t\simeq b^4/v$. 

Within the  RPA approximation  (\call{rpa}),
the inverse osmotic compressibility of the neutral solution
is $ 1/(Nc) + v +w^2 c$. That of the charged solution is 
$ 1/(Nc) + v +w^2 c + 4\pi \lb f^2 /\kappa^2$. 
Hence, in the small wavevector limit,
the electrostatic interactions give rise to 
an additional excluded volume $v_{el}= 4\pi \lb f^2 /\kappa^2$. 

At high wavevectors, the chains are gaussian and the structure 
factor decays as $S(q)\simeq 12/(q^2b^2)$. At small salt concentrations,
the structure factor given by (\call{rpa}) has a peak at a wavevector $q^*$ given by
$$\eqalign{
(q^{*2}+\kappa^2)^2={ 24\pi \lb f^2 c \over {g'b^2}}
}\eqno(peak)
$$
where $g'$ is the derivative of $g$ with respect to $(Nq^2b^2/6)$, equal 
to $1/2$ when $(Nq^{*2}b^2/6)$ is large and to $1/3$ when it is small.
The peak position is independent of the solvent quality. 
In the absence of salt, 
it scales as $q^* \simeq f^{1/2} c^{1/4}$ 
and crosses over smoothly to the peak position in the semidilute range $q^*\simeq 1/\xi$ given by 
equation (\call{xi}) at the concentration $c^{**}$. As the salt concentration is increased, 
the peak shifts towards zero wavevector and disappears when 
$\kappa^4 > ( 24\pi \lb f^2 c )/ (g'b^2)$. 
At higher ionic strength the polymer solution behaves as a 
neutral polymer solution and the structure factor
 decreases monotonically with the wavevector. 
The structure factor given by equation (\call{rpa}) is in good agreement with
 neutron scattering experiments on weakly charged polyacrylic acid or 
polymethacrylic acid solutions in water \refto{Candau2}.

The osmotic pressure of the solution can also be calculated from the RPA (one loop) approximation. 
In the absence of salt it is equal to
$$\eqalign{
{\pi \over k_BT}= fc + {1\over 2} vc^2 + {1\over 3} w^2 c^3 - A f^{3/2} c^{3/4} \lb^{3/4} b^{-3/2}
}\eqno(rpapres)
$$
where $A$ is a numerical constant of order unity \refto{BE}. 
The first term is the osmotic pressure of the counterions 
discussed extensively above; 
the two following terms are due to the non electrostatic 
interactions between monomers.
The last term is due to the charge fluctuations and is very similar to the 
so-called polarisation pressure of simple electrolytes.

Finally, the cooperative diffusion constant of the solution can be calculated from the structure
factor using linear response theory and describing 
the hydrodynamic interactions 
by the so-called Oseen tensor \refto{AJL}. 
When the electrostatic interactions dominate 
over the excluded volume interactions, the cooperative 
diffusion constant is given by
$$\eqalign{
D={k_BT \over {6\pi \eta}}{\alpha^3 \over \kappa^2} 
(1+\kappa^2 /\alpha^2) (2+\kappa^2 /\alpha^2)^{-1/2}
}\eqno(rpadiff)
$$
where $\eta$ is the solvent viscosity and where the wave vector $\alpha$ is defined as 
$\alpha^4= (q^{*2}+\kappa^2)^2$. In a first approximation, 
the cooperative diffusion constant varies
as $D\simeq f^{3/2}c^{3/4}(fc+2n)^{-1}$. It is thus a non monotonic 
function of the monomer concentration $c$.
It increases at low concentration, reaches a maximum when the counterion concentration 
is of the order of the salt concentration and
 decreases at higher concentration. 
The unusual decrease at high concentration is due to the coupling to the counterion 
gas which dominates the compressibility of the solution. 
This might explain the unusual variation of the diffusion constant observed 
on polyacrylic acid solutions and gels.

A summary of the various regimes expected for polyelectrolyte solutions 
in the semidilute and concentrated ranges is given in figure 2.

{\it 6.5 Mesophase formation in poor solvents}

In a poor solvent below the $\theta$ temperature, the polyelectrolyte 
chains are subject to two antagonist forces,
 the attractive (Van der Waals like) 
interaction and the repulsive Coulombic interaction. As explained for 
isolated chains in section 2, the attractive interaction is dominant at short length scales and 
induces a local collapse of the solution. If the salt concentration 
is not too high, phase separation 
is not expected at a macroscopic scale but at a mesoscopic scale. The polymer 
solution is expected to form mesophases where the polyelectrolyte concentration 
is not homogeneous and where polymer dense and polymer dilute regions are periodically arranged.
These phases are very similar to the mesophases observed in surfactant or
 block copolymer systems and various symmetries 
are possible (lamellar, cubic and hexagonal phases) \refto{LE80,BF90}.
Qualitatively, the mesophases are stabilised by 
the translational entropy of the counterions. If a 
macroscopic phase separation occurs, the  separated phases must be electrically neutral and 
the translational entropy of the counterions is low. When a mesophase is formed, 
the electroneutrality can be locally violated, this is associated to a cost in electrostatic 
energy but the translational entropy of the counterions is tremendously increased compared to a neutral dense state.

In the vicinity of the $\theta$ temperature, mesophase formation 
can be investigated using the same random phase approximation as in the previous section. 
The structure factor is given by equation (\call{rpa}) 
where the excluded volume $v=-b^3\tau$ is now negative, $\tau$ being the temperature shift from
the $\theta$ point. The structure factor has a strong peak at a wavevector $q^*$ 
and this peak diverges as the temperature is lowered ($v$ becoming more negative) indicating the 
formation of a periodic structure with a period $2 \pi /q^*$.
In the absence of salt, the spinodal for mesophase formation obtained from 
the divergence of the peak of the structure factor occurs when
$$\eqalign{
{1\over N} +vc + w^2 c^2 + {1\over N} 
g(Nq^{*2}b^2/6) + {{g'b^2}\over 6}\left({ 24\pi \lb f^2 c \over {g'b^2}}\right)^{1/2}=0
}\eqno(mesop)
$$
The qualitative shape of this spinodal line in 
a temperature $\tau$-concentration $c$ plane is shown in figure 3. 
The temperature and the concentration at the maximum of the spinodal line are given by 
$\tau_m\simeq f^{2/3} (\lb /b)^{1/3}$ and $c_m \simeq b^{-3} f^{2/3} (\lb /b)^{1/3}$. 
Upon addition of salt, the period  of the mesophase increases and when $q^*=0$ 
($\kappa^4= ( 24\pi \lb f^2 c)/( g'b^2)$), 
the mesophase transition becomes a macroscopic phase separation very similar 
to the demixing transition of a polymer in a poor solvent.

In this weak segregation limit the symmetry of 
the phases and the topology of the phase diagram have been calculated 
by the Moscow group \refto{dormidontova}
by studying the non linear fluctuations of the solution.

At a temperature much smaller than the $\theta$ temperature, 
in the strong segregation regime, the  interface between 
the dilute and dense polymer regions is sharp and can be described in terms of a surface tension. 
In a very dilute solution micelles are expected to form and 
their characteristics have been studied in references \refto{JL91,Markorabin}. 
At higher concentrations periodic phases are 
predicted. A partial phase diagram considering only the lamellar ordering 
has been determined by Nyrkova et al. \refto{nyrkova2} by a direct minimisation of the mean field free energy.

Most experimental results seem to be obtained outside the region of 
the phase diagram where mesophases form. To our knowledge, mesophase 
formation has been reported only in one case \refto{tanakahanshibayama} 
and no detailed 
study of the mesophase structure seems to exist.

\smallskip

{\bf 7. Dynamical properties}

\taghead{7.}

In this section, we briefly discuss some dynamic properties of polyelectrolyte solutions. 
Only two quantities are considered: the mobility and the electrophoretic 
mobility of a chain in a very dilute solution and the viscosity of the solution. 
In an electrophoresis experiment, the hydrodynamic interactions are screened 
and the electrophoretic mobility is independent of the molecular weight. The separation of 
molecules of different size (such as DNA) is thus not possible by simple electrophoresis. 
The concentration variation of the viscosity of a polyelectrolyte
 solution is very different from that of an organic neutral polymer solution. 
This effect has been known experimentally for a long time 
(it is sometimes referred to as the polyelectrolyte effect) but it remains poorly understood.

{\it 7.1 Mobility and electrophoretic mobility of a single charged chain}

The theory of transport properties in dilute electrolyte solution has been
worked out long ago by Debye, Onsager and Falkenhagen (see 
\refto{RE68},\refto{MA81}, and historical references therein). The existence 
of long range Coulomb interactions between the species in solution results
in a deviation of the transport coefficients from the pure solvent values
 (ionic mobility, conductivity,
electrophoretic mobility and viscosity) that increases as the square root of 
the concentration, rather than the linear increase which is the rule
for neutral species. The origin of this anomalous behaviour can be traced 
back to two effects known as the relaxation effect and the electrophoretic 
effects. Although these effects have different physical origins, they both produce 
a variation of the transport
coefficients
as the square root of the ionic concentration.

The relaxation effect can be described as follows. A test charge
placed in an ionic solution is surrounded by a spherically symmetric
polarisation cloud where charges of oppposite sign are predominant, 
that can be described by the standard Debye-H\"uckel theory. 
If the test charge moves at a constant velocity ${\bf U}$ with respect to the solvent,
 the polarisation cloud is slightly distorted, 
and the charge experiences 
an additional electric field that tends to  slow down 
its motion. The resulting
 force therefore appears as an electrostatic 
 friction on the
test charge. 
More generally, the moving charge $Q$ (position $\RR$, velocity ${\bf U}$)
 in an electrolyte solution 
can be shown (see Appendix B)  to produce an electric field inside the solution
of the form
$$\eqalign{
 {\bf E}(\rr ) = - \TT(\rr -\RR) {\bf U}
}\eqno(erelax)
$$
where the $\alpha$  $\beta$ component of the matrix $\TT$ is
$$\eqalign{
T_{\alpha \beta}(\RR) = 
 {1\over (2\pi)^3} \int d^3{\bf k}\ \exp(i{\bf k}.{\bf R})
{k_{\alpha} k_\beta \over k^2} {Q \zeta_0 \kappa^2 \over
 k_B T (k^2+\kappa^2)^2}
}\eqno(T)
$$
The ionic mobility $\zeta_0^{-1}$ is for simplicity 
assumed to be identical for all small ions.

The electrophoretic effect is important only for the description
 of ionic motion
driven by an external electric field (electrophoresis).
 In that case,
the force that drives a charge motion 
also acts on the surrounding
polarisation cloud. The solvent velocity field that results from this 
body force acting on the polarisation cloud modifies the relative
velocity of the charge with respect to the solvent  creating an increased 
dissipation. In other words, the volume of solvent 
within a distance $\kappa^{-1}$ of the moving charge is
dragged along as the charge moves, with a friction coefficient
of order $\eta \kappa^{-1}$, where $\eta$ is the solvent viscosity.
In the language used to study polymer dynamics, the electrophoretic effect 
can be described as a screening of the hydrodynamic interactions
{\it when the motion is created by an electric field}.
The hydrodynamic interactions are usually  described in terms of
the Oseen tensor $\HH(\rr)$ \refto{DE86}, which gives the velocity 
field that results from the application of a local force $\FF_0 \delta(\rr)$ localized
at the origin. When the force results from an electric field $\EE_0$ applied 
to a charge $Q$ located at the origin, it is easily shown (see Appendix B)
that, if the polarisation cloud is described
within the Debye-H\"uckel approximation,  the velocity field is given by
$$\eqalign{
\exp(-\kappa r) \HH(\rr) Q\EE_0
}\eqno(osscr)
$$
This velocity field is screened in a similar way as the electrostatic interaction and decays 
as $\exp(-\kappa r)/r$. The 
hydrodynamic interaction with neighboring particles is also
screened over a distance 
$\kappa^{-1}$.

The extension of these concepts to macroions has been the subject of
numerous contributions. Much of this work however is concerned 
with spherical (\refto{RS89}, and references therein)
or rodlike \refto{SH82} particles. The case of flexible polymers
has been considered by Hermans \refto{Hermans}
Manning \refto{MA81} and Muthukumar \refto{MU94}.
In reference \refto{MA81}, the relaxation and electrophoretic effects
are treated on a different footing. 
In reference \refto{MU94}, the relaxation
effect is ignored. In the following, we present a simplified description
that allows for a simultaneous treatment of the two effects at a microscopic
level. Our conclusions are essentially similar
\note{ The result that is arrived at in \refto{MA81}
can be obtained within our analysis by ignoring the second term
in the r.h.s. of equation (7.6). The result that is obtained for the
electrophoretic mobility is then $\mu_e =f e  (H_{s0})
 / (1+ fe (H_{s0}) T_0 )$, which is only slightly different
 from (7.9), since $(H_{s0})$ is independent of molecular weight.
A problem would arise, however, if the same approximation were made
in the calculation of the mobility.}
of \refto{MA81}. They  differ from those of reference \refto{MU94}, which
we believe to be in error.

We consider here two types of experiments. In a sedimentation experiment, 
the external force is the same on all monomers and has no effects on the small ions. 
The chain mobility $\mu$ relates the velocity of the center of mass of the molecule $\bf v_G$
to the total force acting on the molecule $\bf F$: ${\bf v_G}= \mu \bf F$. 
In the electrophoresis experiment, the external 
force is due to the applied electric field $\bf E$
acting both on the charged monomers and on the free small ions. 
The electrophoretic mobility is defined 
by ${\bf v_G}= \mu_e \bf E$.

The model that we use to describe the dynamics of the 
charged polymer is a simple extension 
of the standard Zimm model of neutral polymers \refto{DE86}.
The velocity and position of monomer $i$ are denoted respectively 
by $\RR_i$ and $\vv_i$, the velocity field of the solvent is $\uu(r)$. 
The equation of motion of monomer $i$, bearing a charge $fe$ results 
from a balance between a friction force and all the other forces acting on the monomer:
$$\eqalign{
\zeta ({\bf v_i} - {\bf u}({\bf R}_i) ) =
 {\bf F}_i + f e {\bf E_r}({\bf R}_i)
}\eqno(vi)
$$
The monomer solvent friction coefficient is $\zeta$. The force
 $\FF_i$ is the sum of an external force $F_{ext}$ 
(the sedimentation force or the applied electric field), a Langevin random force
${\bf \theta}_i$
and of an intramolecular force $\FF_{pol,i}$.  The last term $\EE_r$ 
is the electric relaxation field.

The velocity field of the solvent is given by an 
Oseen-type relation
$$\eqalign{
{\bf u}({\bf R}_i) = \sum_j \HH(R_i-R_j) [ s(R_i-R_j) {\bf F}_{ext} 
+\FF_{pol,j}+{\bf \theta}_j] 
+ \HH_s(R_i-R_j) f e {\bf E_r}({\bf R}_j)]
}\eqno(solv)$$
the screening factor $s(r)$ is equal to $\exp(-\kappa r)$ in the case
of electrophoresis and to unity otherwise. If we assume that 
the solvent velocity field varies slowly over a scale $\kappa^{-1}$,
the relaxation field $E_r(\RR_i)$ on monomer $i$ is obtained from 
(\call{erelax}) as
$$\eqalign{
{\bf E_r}(\RR_i) = - \sum_j \TT(\RR_i-\RR_j) ({\bf v}_j- {\bf u}(\RR_j)) \ . }\eqno(er)
$$
These equations can be treated in the spirit of the Kirkwood-Rieseman
preaveraging approximation, by replacing the tensors $\HH(\RR_i-\RR_j)$,
$s(\RR_i-\RR_j) \HH(\RR_i-\RR_j)$ and $\TT(\RR_i-\RR_j)$ 
by their equilibrium average values. Introducing
$$\eqalign{
H_0  =& N^{-1}  <\sum_{ij} \HH(\RR_i-\RR_j)>\cr
H_{s0}=&  N^{-1} <\sum_{ij} \exp(-\kappa |\RR_i-\RR_j|) \HH(\RR_i-\RR_j) >\cr
T_0   =& N^{-1} <\sum_{ij} \TT(R_i-R_j)>
}\eqno(pav)
$$
the mobility $\mu$  and electrophoretic mobility 
$\mu_e$
are expressed as:
$$\eqalign{
\mu  = {1\over N} \left[ H_0 -H_{s0} fe{T_0 \over \zeta} 
(1+ fe {T_0 \over \zeta})^{-1}\right] 
}\eqno(mu)
$$
$$\eqalign{
\mu_e =f e  H_{s0}  (1+ fe {T_0 \over \zeta})^{-1}
}\eqno(mue)
$$
The mobility is given
by the usual Kirkwood-Riseman result \refto{DE86}, with a correction
that arises from the relaxation effect. This relaxation correction
reduces the mobility. Using equation (\call{T}),
$T_0$ can be computed as 
$$\eqalign{
fe {T_0 \over \zeta} = & f^2  \lb  {\zeta_0 \over  3 \zeta}
 \int {{\rm d}^3\qq \over (2\pi)^3} S(q) { \kappa^2 \over (q^2+\kappa^2)^2 } \cr
= & f^2 \kappa \lb  {\zeta_0 \over  3 \zeta} N^{-1} <\sum_{ij}\exp(-\kappa |\RR_i-\RR_j|) >
}\eqno(t0)
$$
where $S(q)$ is the structure factor of the chain (normalized
 so that $S(q=0)=N$). From the second line of (\call{t0}) it is 
seen that the effect of the relaxation field is independent of molecular
weight whatever the chain conformation and is small when the charge fraction $f$ is small. 
For example, if the polymer is rodlike over a distance  
$\kappa^{-1}$, $fe {T_0 / \zeta}= f^2 \lb /(3b)  (\zeta_0 / \zeta)$.
The averaged screened Oseen tensor, 
$H_{s0}$, is also independent of molecular weight and proportional to 
$ - \ln(\kappa b) $. The mobility is thus dominated 
by the Kirkwood Riseman contribution, and is essentially the same as for 
a neutral chain with the same conformation. The relaxation correction becomes 
important for strongly charged rodlike chains in the 
vicinity of Manning condensation threshold. In this case, 
the mobility is reduced by a factor of order unity from the Kirkwood Riseman value.
The approximations that we have made are only valid for
long chains, $R>>\kappa^{-1}$.  In view of the complex  form
of equation (\call{mu}), it is not surprising that experimental
results for polyelectrolyte diffusion constants are often
intermediate between the "non-draining" and "free draining" limits
\refto{Reed}.

The electrophoretic mobility $\mu_e$, on the other hand, 
is in any case independent of molecular weight because 
of the screening of hydrodynamic interactions. For weakly charged polyelectrolytes, 
the relaxation correction is small and the electrophoretic mobility can be rewritten 
in terms of the structure factor of the chains as
$$\eqalign{
\mu_e = {fe \over {3\eta \pi^2}}
 \int {{\rm d}k} S(k) {k^2 \over (k^2+\kappa^2)}
}\eqno(me)
$$
For a rodlike molecule, the electrophoretic mobility decreases 
logarithmically as  the ionic strength is increased 
$\mu_e= -{fe \over {3\eta \pi^2 b}} \ln \kappa b$. For a flexible chain, 
using the stretched chain model of section 2 to describe the local conformation of the chains, 
 we obtain $\mu_e= -({fb/\lb})^{1/3} (e/{\eta b})\ln \kappa \xi_e$,  
the increase with the charge fraction is weaker and the electrophoretic mobility also decreases 
logarithmically with ionic strength.
The general features of these theoretical predictions are 
in reasonable agreement
with experimental observations.  
A detailed comparison between theoretical and experimental
results is given in \refto{MA81}.

{\it 7.2 Viscosity of polyelectrolyte solutions}

The viscosity of polyelectrolyte solutions has been known
very early 
 to exhibit a behaviour qualitatively 
different from that of
neutral polymer solutions (see \refto{TA61}, and references therein).  
  This behaviour was first 
described by Fuoss \refto{fuoss1} using an empirical law, which gives 
the reduced viscosity $\eta_r = (\eta(c) -\eta_s)/\eta_s c$ 
as a function of the polymer  concentration $c$
$$\eqalign{
\eta_r = {A \over 1 + B c^{1/2}}\ .
}\eqno(fuoss)
$$
Detailed studies \refto{KP88} show that 
the constant $A$, obtained by extrapolating the results
to vanishing  concentration,
is proportional to $N^2 f^2$, while the ratio $A/B$ is proportional 
to $Nf$.

It was later realized that this law is satisfactory 
only for high enough concentrations and  in saltfree
solutions, in which case it reduces
to $\eta_r\sim c^{-1/2}$. At low concentration, 
the behaviour of $\eta_r$
is actually nonmonotonic \refto{pouyeteisenberg}.
 The reduced viscosity $\eta_r$ presents a peak at a finite
concentration, of the order of 
the salt concentration. This peak disappears when enough salt is 
added to the solution. A  phenomenological law, recently proposed by 
Cohen, Priel and Rabin \refto{CP88_1}, seems to account well 
for the experimental observations. According to these authors, the 
intrinsic viscosity is given by
$$\eqalign{
\eta_r ={ A_1 c \over \kappa^3}
}\eqno(rabin)
$$
where $\kappa$ is the Debye screening parameter, and $A_1$ is proportional
to the molecular weight. 

Finally, it is important to note that this anomalous behaviour 
of polyelectrolyte solutions can be rationalized at low concentrations
by using the 
method of isoionic dilution \refto{TA61,PH52}. In this method, the 
viscosity is studied as a function of polymer concentration
for a given value of the ionic strength (i.e. as the solution is
 diluted, salt is added in order to keep the ionic strength constant). 
The increase of the reduced viscosity at low $c$ is then linear,
 as for neutral polymers. In neutral polymer solutions, this linear
behaviour is usually represented in the form 
$\eta_r = [\eta] + k_H [\eta]^2 c $, where $[\eta]$ is the 
intrinsic viscosity and $k_H$ is the Huggins constant. 
In polyelectrolyte solutions, $[\eta]$ is large
and varies strongly with the ionic strength $I$. The data of \refto{PH52},
for exemple, can be represented in the form 
$[\eta] = [\eta](I=\infty) + C I^{-1/2}$ \refto{WO78}. 
The slope of the linear 
term, $k_H [\eta]^2$, is also much larger than the corresponding
quantity in neutral polymers. The Huggins constant $k_H$ varies
as  $I^{-1/2}$ \refto{WO78} \note{In fact, the Huggins constant is
not a particularly useful concept in a system 
where interactions that are not of hydrodynamic origin
are expected to be important.}. This indicates the importance of
interactions between chains, and also shows that, as $c$ is increased
a crossover to the "interaction dominated" behaviour described
by (\call{rabin}) rapidly takes place.

In short, three different concentration
regimes can be distinguished in the behaviour of the reduced viscosity as a function of $c$
(figure 4)
\note{This distinction is a simplification of
the description proposed by Wolff \refto{WO78}, and
seems valid at all but the highest concentrations.}.
In the   low concentration region (left side of the peak),
$\eta_r$ is an increasing function of $c$. This region is most usefully 
 characterized by isoionic dilution curves. The peak region corresponds
to comparable salt and counterions concentrations
($f c\simeq n$). Finally, in the more concentrated 
region, $\eta_r$ is a decreasing function of $c$, $\eta_r \sim c^{-y}$
with $y<1$. Obviously, this description is valid
only in solutions where the salt concentration is small enough,
since if $n$ is large the behaviour of usual neutral 
polymers must be recovered.

At present, a comprehensive theory of polyelectrolyte viscosity
is lacking. A number of interpretations, however, have been 
put forward to explain the unusual behaviour of these solutions.
We will distinguish here two types
of models based either on the conformation
of the polymers or on the interactions between
 different chains. 

In conformation based models \refto{GP76,DR86,RE94,RC94}
(which have been formulated for flexible polyelectrolytes),
it is assumed 
 that the viscosity of the polyelectrolyte solution
is only a consequence of the structural 
properties of the solution. The calculation
proceeds by analogy with the well understood case
of neutral polymers. For semidilute solutions, the viscosity 
can be computed by applying the Rouse or reptation theories
to a chain of correlation blobs of size $\xi$, containing $G_\xi$
monomers. 
The correlation length of polyelectrolyte solutions
has a variation with concentration
very different from that of neutral polymer solutions, so that the behaviour
of the viscosity is also qualitatively different.
This idea was first applied by de Gennes and coworkers
\refto{GP76}. These authors interpreted the $[\eta] \sim N  c^{-1/2}$
 behaviour in saltfree semidilute
solutions  given by Fuoss law as corresponding to Rouse dynamics for a chain 
of correlation blobs of size $\xi \sim \kappa^{-1}$ (equation (\call{xi}))
with $G_\xi = g (\xi/\xi_e) $ monomers. 
The viscosity is then given by
$$
\eqalign{
\eta =  & \eta_s { c\over N} \left({N\over G_\xi} \right)^2 \xi^3\cr
     = & N \eta_s c^{1/2} \xi_e^{-3/2} b^{3}
}\eqno(rouse)
$$
  This   analysis was extended by Dobrynin et al to 
the case where salt is added to the solution \refto{DC95}.
In that case, $\xi$ is given by (\call{xiP}), $G_\xi= c\xi^3$
and the viscosity is 
$$\eqalign{
\eta =  N \eta_s   c^{5/4} \kappa^{-3/2} \xi_e^{-9/4} b^{9/2}.
}\eqno(rouse1)
$$
The crossover  between (\call{rouse}) and (\call{rouse1}) takes place when
the mesh size is equal to $\kappa^{-1}$, like the crossover between 
(\call{xi}) and (\call{xiP}). Note that equation (\call{rouse1})
predicts a maximum in the reduced viscosity $\eta_r(c)$ for $fc=n$. 

Rubinstein and coworkers \refto{DC95} also give an empirical justification
for 
the observation that semidilute polyelectrolyte 
solutions obey Rouse dynamics over a broader range of concentrations
above the overlap concentration than neutral polymer solutions.
 In neutral polymers,
it is observed that entanglements become important when 
each chain interacts with about  $10$ other chains \refto{entanglements}.
If this criterion is applied to polyelectrolytes, 
the concentration $c_{rept}$ for the onset of reptation behaviour 
is given by
$$\eqalign{
{c_{rept} \over N} \left( {N \over G_\xi}
 \xi(c_{rept})\right)^{3/2} \simeq 10 . 
}\eqno(crept)
$$
This can be rewritten as $c_{rept} \xi(c_{rept})^3 \simeq N/100$.
The overlap concentration $c^*$, on the other hand, is given by 
$c^* \xi(c^*)^3 \simeq N$. In polyelectrolyte solutions, 
$c\xi^3(c)$ decreases much more slowly with increasing concentration $c$ than in neutral polymer
solutions,
so that the onset of reptation is expected at higher concentrations. 
If equation (\call{xi}) is used, the ratio $c_{rept}/c*$ is of order $10^4$, while
for neutral polymers it is of order $10^2$.

 Another useful way
of computing the viscosity 
 of neutral polymer solutions, which allows  a description of the 
dilute/semidilute crossover, is to write it  in a scaling form
$$\eqalign{
\eta = \eta_s F\left({c\over c^*}\right)
}\eqno(eta)
$$
where $\eta_s$ is the solvent viscosity, 
 $F$ is a scaling function and $c^*$ is the overlap
concentration. The function $F$ can be 
expanded as a power series ($F(x) = 1 + a_1 x + a_2 x^2..$)
for small values of its argument, i.e. in dilute solutions.
In the semi-dilute regime, the functional form of $F$ depends
on the mechanism (Rouse or reptation) that governs 
stress relaxation.
In this formulation, the specificity of polyelectrolyte solutions
is related to the fact that the interactions between monomers
depend on the polymer concentration. Therefore $c^*$
in (\call{eta}) must now be interpreted as the overlap 
concentration corresponding to a polymer having the same interactions
as the charged polymer at concentration $c$.  For exemple, when the mesh
size is larger than $\kappa^{-1}$, we can write
$$\eqalign{
c^*(c)= N/R(c)^3 .
}\eqno(cstar)$$
Where $R(c)= \kappa^{-1} ({N \over g }\xi_e \kappa)^{3/5}$,
given by  equation (\call{RP}),
is the  radius of an isolated chain
with this interaction.
The viscosity is obtained by writing
$\eta = \eta_s F(c/c^*(c))$, and assuming Rouse behaviour 
$\eta \sim N$. This gives
$$\eqalign{
\eta = \eta_s \left({c\over c^*(c)}\right)^{5/4} 
= N \eta_s c^{5/4} \kappa^{-3/2}
 (\xi_e)^{9/4} b^{-9/2}
}\eqno(rouse2)
$$
which is identical to (\call{rouse1}).

Another "conformational" interpretation
of polyelectrolyte viscosity was given by Reed \refto{RE94}, using 
similar arguments.
His calculation assumes that $c$ is in the vicinity of $c^*(c)$,
so that the $x^2$ contribution in the function
$F(x)$ dominates. The formula used for $c^*$ 
is appropriate for short chains
or large $\kappa^{-1}$. It corresponds to an ideal
behaviour for a chain with a persistence length $\kappa^{-1}$
$$\eqalign{
R(c)^2  = 2  (N/g)\xi_e \kappa^{-1}.
}\eqno(reed0)
$$
This leads to a viscosity 
$$\eqalign{
\eta \sim \eta_s (c/c^*)^2 \sim \eta_s N {c^2 \over \kappa^3} (\xi_e/g)^3.
}\eqno(etareed)
$$
(Note that the same result
would be obtained by requiring Rouse behaviour for 
the viscosity, i.e. $F(c/c^*)\sim N$)
Again, the behaviour $\eta_r \sim f N c^{-1/2}$
is obtained in the absence of salt. With added salt,
the experimentally observed behaviour (\call{rabin})
is reproduced. The main  difference between the interpretations of
Reed and of Dobrynin et al. is the expression used
for $R(c)$. The crossover from (\call{etareed})
to   (\call{rouse1}) can be expected to take
place as $c$ increases and $R(c)$ crosses over from the
ideal expression (\call{reed0}) to the 
Flory expression (\call{R1}). 

Finally, we note that in all
conformation based models the $c^{-1/2}$ behaviour
of $\eta_r$,
corresponds to situations where the role of added
salt is negligible. It is interesting that the value of the
constant $A$ in Fuoss law (\call{fuoss}), which corresponds to a
(generally incorrect) extrapolation of this behaviour
to $c=0$, yields a value $A\sim N/R^3 \sim N^2  f^2$ that corresponds
to the intrinsic viscosity for a saltfree, dilute solution 
of  polyelectrolytes ($R\sim Nf^{2/3}$). This coincidence,
however, is not explained theoretically. At high concentrations,
entanglements become important, and the reduced viscosity
increases with concentration \refto{WO78,DC95}

The conformation based models
appear to provide a qualitatively correct description
of flexible polyelectrolyte viscosity (see \refto{RE94,DC95} for a 
comparison with experimental data).
They only apply, however, to sufficiently concentrated solutions,
near or above overlap concentration.
 The "polyelectrolyte
behaviour" described by equations (\call{fuoss}) or (\call{rabin})
is also observed in solutions of short chains,
or even for charged spherical  particles \refto{Ise}. 
In \refto{CP88_2}, it was suggested that this behaviour could
be described on the basis of the liquid
state "mode-coupling" theories developped for the study of spherical
charged  collo\"{\i}ds \refto{HK}. Unfortunately, these theories
are analytically tractable only in the limit of weak electrostatic coupling
between the particles, which is certainly not realized
in dilute polyelectrolyte solutions (see section 6.1). 
In this weak coupling regime, it was shown
that the viscosity of a suspension of spherical charged
particles (charge $Z$, concentration $c_0$, hydrodynamic radius $R$) is given by 
$$\eqalign{
\eta \sim  \eta_s R c_0^2 Z^4 \lb^2 \kappa^{-3}
}\eqno(HK)
$$
The authors of reference \refto{CP88_2} noticed that this behaviour
is similar to the experimentally observed one (\call{rabin}), if the 
polymer chain is assumed to bear an effective charge
$Z^*= (fN)^{1/2}$. The particle concentration is replaced by
$c/N$, and the hydrodynamic radius is 
proportional to $N$.
The physical origin of the effective charge $Z^*$, however,
is unclear.

This discussion indicates that a full theoretical understanding
of the viscosity of polyelectrolyte solutions is far from being achieved.
The main results can be summarized as follows. In all models, a maximum
in $\eta_r$ is predicted when the salt and the counterion
concentration are of the same order of magnitude, $fc\simeq n $. The maximum 
can occur, depending on chain length and salt concentrations,
either in the dilute or
in the semi-dilute region. For short chains or small salt concentrations,
it thus takes place in dilute solution. The appropriate theory
 to describe the peak should be similar to that of references
\refto{HK,CP88_2}.
At concentration higher than the peak concentration,
the reduced viscosity behaves as $fNc^{-1/2}$. Although
the behaviour is similar above and below the overlap concentration, 
the interpretations that have been proposed are very different.
For longer chains, or higher salt concentrations, $fc \simeq n$ corresponds
to a semidilute solution. In that case, the viscosity  can be described
using equations (\call{rouse1}) (for $c$ larger than $c^*$)
or (\call{etareed}) (for $c$ near $c^*$). Below $c^*$, the 
appropriate description is
again that of \refto{HK,CP88_2}.

\smallskip

{\bf 8. Conclusions}

In this review we have presented the recent theories that describe the
conformational and dynamical properties of charged polymer chains.
Much of the focus has been put on the long range character of the electrostatic interactions
and on the role of the small counterions that insure the electrical neutrality. 
Little attention has been paid to interactions of non electrostatic origin
\refto{OD94} which 
have only been introduced at some places in terms of virial coefficients. Most experiments
on polyelectrolyte solutions are made in water that is not a good solvent for 
polyelectrolytes which are organic polymers. The non electrostatic 
interactions between molecules dissolved in water are not simple Van der Waals interactions and 
have components due to hydrogen bonding or to the hydrophobic 
effect \refto{IS85}. It is not obvious that these 
very specific interactions can be modelled simply in terms of virial coefficients. Experimentally
these strong attractive interactions lead in some cases to the formation 
of aggregates which can be responsible for the slow modes 
\refto{Cedlak,Reed,Schmidt} sometimes observed when measuring 
the relaxation of concentration fluctuations via quasielastic light scattering. 
The simple theoretical 
models that we have described do not take into account 
these complicated effects 
which are in many cases dominant. This makes a direct comparison between 
the theory and the experimental results rather difficult and we have not attempted here any 
quantitative comparison.

Despite this difficulty, some experimental features seem 
to be rather general in polyelectrolyte solutions. The structure factor of a 
polyelectrolyte  solution always shows a peak at low ionic strength. The position of 
the peak has been discussed insection 6. both in dilute and semidilute solutions.
Most experiments are also consistent with a rodlike behavior of 
the polyelectrolyte chains at small length scales. The rigidity of the chain is then 
characterized by an electrostatic persistence length. The variation of the persistence 
length with ionic strength or with the polymer concentration 
is however still a matter of controversy as seen in section 4. 
and more complete theories are certainly needed. 
The effect of heterogeneities of the charge distribution or of non electrostatic 
interactions on the persistence length do not seem to have been discussed. A good agreement 
between theory and experiment seems to exist only for rigid polymers where 
the theory of Odijk or Skolnick and Fixman \refto{MW83,Hagerman} gives a quantitative description 
when counterion condensation is properly taken into account. 
Fuoss law and the variation of the viscosity of polyelectrolyte solutions with the concentration
have also been  extensively confirmed experimentally. 
The theoretical results presented in section 7.  remain at the level 
of scaling laws and are not based on very 
systematic arguments. It should be also noted that the derivation of Fuoss law 
in semidilute solution based on the Rouse model
implicitly assumes that there is only one length scale and therefore that 
the chain persistence length is of the order of the correlation length $\xi$
as suggested by some of the theoretical models. 

An alternative way to study the properties of polyelectrolyte solutions is numerical simulations \refto{Kremerdunweg}. 
All the parameters are well controlled in the simulations 
and they  should allow to sort out the respective 
roles of the electrostatic and non electrostatic interactions and thus 
to provide quantitative tests of the theoretical models. Many simulations have been performed on  
isolated polyelectrolyte chains in a salt free solution. There is in general quantitative 
agreement between the results and the electrostatic blob model. The simulations 
in the presence of salt often performed by using the Debye-H\"uckel potential are less conclusive. 
For rigid polyelectrolytes, the agreement with the Odijk-Skolnick and Fixman theory is good; 
for flexible weakly charged polyelectrolytes, there is no good separation of length scales between 
the screening length and the persistence length and a quantitative test of the theories is difficult.
Recently, Kremer and Stevens \refto{KS94} have performed simulations on semidilute polyelectrolyte solutions, taking into account
explicitely the discrete counterions. 
The polymers are strongly charged flexible chains in the vicinity of Manning 
condensation threshold. This is a very heavy 
numerical work but it should allow a very detailed description of the influence of 
electrostatic interactions in polyelectrolyte solutions. 
The extension of these simulations to weakly charged or more 
rigid polyelectrolytes for which the analytical theories have been constructed would 
provide strong tests of these theories.

The review has also been limited to simple polyelectrolyte 
chains comprising only monomers with the same charge and neutral monomers. There are 
several other polymeric sytems where electrostatic interactions play an important role. 
Polyampholytes \refto{HiggsJoanny,KardarKantor}
are polymers that carry charges of both signs. Due to the attractions between opposite charges, 
they are often insoluble in salt free water and can only be dissolved at a finite ionic strength. 
They also show an antipolyelectrolyte effect, the reduced viscosity increasing with ionic strength.
Polysoaps are comblike polymers, often polyelectrolytes, where hydrophobic 
side chains are grafted on the backbone \refto{TA61,Turnerjoanny}. 
They have properties combining those of polyelectrolyte chains 
and of small ionic surfactants and form 
for example intramolecular micelles.  Another system of interest is mixtures between 
neutral and charged polymers \refto{nyrkova3,leibleriliopoulos}
 which have been studied both experimentally and theoretically. 
The addition of a charged polymers can increase the solubility of a neutral polymer in water. 
The formation of mesophases is also expected in these systems.

Throughout this review we also have implicitly assumed that the solvent
(essentially water) is a continuous dielectric medium with a uniform dielectric constant.
At high polymer concentration the dielectric constant crosses over from that of the polar 
solvent to that of an organic medium (essentially polymer) \refto{khokhlov1}. If the dielectric constant is low, 
charges of opposite signs are not dissociated and the polymer has an ionomer behavior dominated 
by attractive interactions between dipoles. Even in a more dilute solution, 
it is not clear that locally,
the dielectric constant has the solvent value; this could turn out 
important in Manning condensation theory for example. 
There seems to be little theoretical work on these issues. Also, the small ions
have been considered as pointlike, all specific interactions due
to hydration or to the finite size of the ions have been neglected.
Small differences between ions due to these effects 
are usually observed.

One of the essential conclusions of this review is that polyelectrolytes 
remain a poorly understood state of matter and that much work is still 
needed both from the experimental and from the theoretical point of view to reach 
a degree of understanding equivalent to that of neutral polymers. 
>From a theoretical viewpoint, polyelectrolytes are in most cases
strongly coupled systems, involving many different length scales.
It might well be that in such situations,  an appropriate
description of the system can only be obtained by the use of 
sophisticated liquid state theories, such as those recently developped 
for dense neutral polymers \refto{CS}. Experimentally,
it is surprising that relatively few systematic studies
on charged polymers, 
using modern investigation tools, in particular neutron scattering,
are available.
Such studies were initiated in the late seventies (see e.g \refto{Jannink}),
but have not been as conclusive as the equivalent studies 
on neutral polymers. Finally, it is obvious
that the area of polyelectrolyte solutions ofters a number 
of problems that can be approached by simulation. Simulations that explicitly include
counterions can be
used to assess the accuracy of the Debye-H\"uckel or Poisson-Boltzmann
approximations \refto{woodward}. 
Density functional based simulation methods
\refto{Madden} for classical charged systems can be used to go 
beyond  such  approximations, even in the presence of added 
salt. Finally, the dynamical properties of charged polymers
have not, up to now, been investigated in numerical studies.

\endpage

\centerline {\bf APPENDIX A}
\centerline {\bf Effective interaction between charged monomers}

\taghead{A.}

In this appendix, it is shown  that 
the effective interaction between charged monomers can, within the linear 
response approximation, rigorously be written as a sum of pairwise 
additive interactions (equation (\call{vdh}). As a byproduct
 of the calculation,
  the various stucture factors 
(ion-ion, polymer-polymer and polymer-ion) are linked
within this approximation,
by simple relations. 

The system is a  polymer solution with a concentration
$c$ of (positively) charged monomers, and a concentration $n$ of monovalent 
added salt.
For simplicity we assume that the couterions are identical 
to the salt cations. 
The densities of charged monomers, charged positive and negative ions
are denoted by $\rho_P(\rr)$, $n_+(\rr)$ and $n_-(\rr)$, respectively.
 By integrating out
 the coordinates of
the small ions, the partition function 
of the system can formally be written as
$$\eqalign{
Z= {\rm Tr}_{_P} \exp(- (H_P + F_i[\psi])/k_BT
}\eqno(Z)
$$
where ${\rm Tr}_{_P}$ denotes an integration over monomer coordinates,
 and $H_P$ is the energy of the polymer (\call{H}). $ F_i[\psi]$
 is the free energy of the small ions in a potential that is the 
sum of the external potential and of the electrostatic potential
$\psi(\rr)$ created by the polymer
$$\eqalign{
\psi(\rr)= k_B T \lb \int d\rr' {\rho_P(\rr') \over |\rr-\rr'| } \ .
}\eqno(psi)
$$
 The free energy  $F_i$ is obtained by minimizing with respect to the ionic densities
$n_+(\rr)$ and $n_-(\rr)$ a free energy functional, which in the
mean field or Poisson-Boltzmann approximation reads
$$\eqalign{
 & k_BT \int d\rr (n_+(\rr) \ln(n_+(\rr)) + n_-(\rr) \ln(n_-(\rr)) )\cr 
& + {1 \over 2 } \lb K_B T \int d\rr d\rr' 
{(n_+(\rr)-n_-(\rr)) (n_+(\rr')-n_-(\rr')) \over  |\rr-\rr'| }
+ \int d\rr (n_+(\rr)-n_-(\rr))\psi(\rr)}\eqno(fro)
$$
The linear response, or Debye-H\"uckel, approximation, involves a further 
simplification of equation (\call{fro}), by expanding the
first two terms to second order in the deviations of $n_+$ and 
$n_-$ from their average values. The minimisation is easily
carried out, and the resulting expression
for $F_i$ is 
$$\eqalign{
F_i[\psi] = - \int d^3\kk  {1\over 2} {4\pi\lb \over k^2} 
{\kappa^2 \over k^2 +\kappa^2} 
\rho_P(\kk) \rho_P(-\kk)  \ .
}\eqno(fmin)
$$
with $\kappa^2 =4 \pi \lb (2 c + n )$.
The interaction that results 
from adding  this contribution from the ionic {\it free energy}  to the
electrostatic repulsion between the monomers, 
$\int d^3\kk  {1\over 2} {4\pi\lb \over k^2}\rho_P(\kk) \rho_P(-\kk) $,
 gives the screened pair potential (\call{vdh}).

The Debye-H\"uckel calculation can easily be extended to compute the 
response of various densities to external fields, or equivalently 
the partial structure factors. The results are simple
in the case of a salt concentration  much larger than
 the counterion concentration ($n >> c$). Defining the ionic charge 
density as $\rho_Z = n_+ - n_-$, the following relationships are obtained
between the Fourier transforms of the polymer-polymer , polymer-charge and charge-charge 
correlation functions 
 (denoted by $S_{PP}, S_{PZ}$ and $S_{ZZ}$, respectively)
$$\eqalign{
S_{PZ}(k) & = - {\kappa^2 \over k^2 +\kappa^2 } S_{PP}(k) \cr
S_{ZZ}(k) & = 2n {k^2 \over k^2 +\kappa^2 } + 
\left({\kappa^2 \over k^2 +\kappa^2 }\right)^2 S_{PP}(k) \ .
}\eqno(s)
$$
The total charge structure factor,
$S_{PP}+ S_{ZZ} - 2 S_{PZ}$, vanishes as $2 n k^2/\kappa^2$ in the
small $k$ limit, as implied by the Stillinger and Lovett sum rules \refto{HM86}.

\endpage
\centerline{\bf APPENDIX B}
\centerline{\bf Relaxation and electrophoretic effects}
\taghead{B.}

In this appendix, we derive equations  (\call{erelax}) and 
(\call{osscr}), describing respectively the 
relaxation and electrophoretic effects. As usual, the two effects 
are considered independently \refto{RE68}, since the coupling between them
is relevant only at a level that goes beyond
the limits of the Debye-H\"uckel linearized theory.

{\it B.1 Relaxation field}

The relaxation field is the electric field 
created by  a charged particle (charge $Qe$) moving in a 
monovalent salt solution (concentration $n$)
with a constant velocity $\VV$. As explained in section 7.1,
this relaxation field arises from the deformation of the
Debye-H\"uckel polarisation cloud when the charge
is set into motion. At equilibrium, the densities $n_+(r)$
of positive
and $n_(r)$  of negative ions that surround the charged particle
(placed at the origin) are
$$\eqalign{
n_\pm^{(eq)}(r) = n (1 \mp Q {\lb \over r} \exp(-\kappa r)) \ .
}\eqno(req)
$$
In a time dependent situation, however, the Boltzmann statistics that is
used to obtain these densities must be replaced with a diffusion equation
$$\eqalign{
\zeta_0 {\partial n_\pm (\rr,t) \over \partial t}
= \nabla \cdot \left( k_B T 
 \nabla
n_\pm(\rr,t) \mp n_\pm(\rr,t) \EE(\rr,t) \right)\ .
}\eqno(smol)
$$
where $\zeta_0$ is the ionic mobility.
The electric field $\EE$ is the sum of the field $\EE_P(\rr-\VV  t)$ created
by the moving particle and of the electric field created by the ionic
charge density, $\EE_i$, which verifies Poisson equation 
$$\eqalign{
 \nabla \cdot \EE_i(\rr,t) =
4\pi \lb (  n_+(\rr,t)-n_-(\rr,t)) \ .
}\eqno(pois)
$$
 The solution 
of (\call{smol}) in the stationnary state is
of the form $n_\pm(\rr,t) = n_\pm^{(eq)}(\rr-\VV t) + 
\delta n_\pm(\rr-\VV t)$. Inserting this form into (\call{smol}) 
and (\call{pois}),
and linearizing with respect to $\VV$ and $Q$, the following 
equation for $\delta n_+ - \delta n_-$ is obtained.
$$\eqalign{
 (\nabla^2 +\kappa^2)(\delta n_+(\rr) - 
\delta n_-(\rr))
= {\zeta_0 \over k_BT }\VV \cdot \nabla \left[
(2 n Q {\lb \over r} \exp(-\kappa r) \right] \ .
}\eqno(relax1)
$$
The resulting ionic charge density can be expressed as
$$\eqalign{
(\delta n_+(\rr)- \delta n_-(\rr) (\rr)
= {1\over (2 \pi)^3} \int d^3\kk\ \left( {\zeta_0 Q e \kappa^2 
\over k_BT}\right) {i \kk\cdot\VV \over (k^2+\kappa^2)^2 }\exp(-i\kk\cdot\rr)
}\eqno(relax2)
$$
and the resulting electric field is given by equation (\call{erelax}).

\smallskip

{\it B.2  Electrophoretic effect}

\smallskip

We now consider the calculation of the hydrodynamic
velocity field around a charge $Q$ at the origin, when the charge
experiences a force $Q\EE_{ext}$ created by an external
 electric field $\EE_{ext}$. The solvent 
experiences a force $Q\EE_{ext} \delta(\rr)$ 
created by the charge at the origin,
and a body force $e (n_+(\rr)-n_-(\rr))\EE_{ext}$. The  equation
that determines the hydrodynamic velocity field $\UU(\rr)$ in the solvent is
$$\eqalign{
 - \eta_s \nabla^2 \UU + \nabla P = Q\EE_{ext} \delta(\rr)
+ e (n_+(\rr)-n_-(\rr))\EE_{ext}\ ,
}\eqno(stokes)
$$
where $\eta_s$ is the viscosity and $P$ the pressure. $(n_+(\rr)-n_-(\rr))$ 
is the charge density within the
 {\it equilibrium}
Debye-H\"uckel
polarisation cloud. In Fourier
space, the velocity field is given by a generalization of Oseen's
formula,
$$
\eqalign{
U_i(\kk) = {1\over \eta_s (k^2+\kappa^2) } 
(\delta_{ij} - k_ik_j /k^2) (Q \EE_j) .
}\eqno(oseen)
$$
Which gives in real space,
$$\eqalign{
U(\rr) = \HH_s(\rr) (Q\EE_{ext})
}\eqno(U)
$$
with the screened Oseen tensor $H_s$ being
$$\eqalign{
H_{s,ij}= {\exp(-\kappa r) \over 8 \pi\eta_s r }( \delta_{ij} + r_ir_j/r^2) .
}\eqno(hs)
$$

\refis{AJL} A. Ajdari, L. Leibler, J.-F. Joanny,
{J. Chem. Phys.}, {\bf 95}, 4580 (1991)

\refis{BB93} J.-L. Barrat and D. Boyer, {\it J. de Physique II},
{\bf 3}, 343 (1993).

\refis{BJ93} J.-L. Barrat and J.F. Joanny, {\it Europhysics Letters},
{\bf 24}, 333  (1993).

 \refis{BP92} J-P. Bouchaud, M. M\'ezard, G. Parisi and J.S. Yedidia,
{\it J. Phys. A} {\bf 24}, L1025 (1992)

\refis{BD94} D. Bratko and K. Dawson, {\it Macromolecular Theory and Simulation}.
 {\bf 3}, 79 (1994); {\it J. Chem. Phys.}, {\bf 99}, 5732 (1993).

\refis{BR47} H.C. Brinkman, Appl. Sci. Res., {\bf A1}, 27 (1947)

\refis{LB82} M. Le Bret, {\it J. Chem. Phys.} {\bf 76}, 6243 (1982).


\refis{CP88_1} Y. Rabin, J. Cohen and Z. Priel,
{\it J. Pol. Sci C}, {\bf 26}, 397 (1988).

\refis{CP88_2} J. Cohen, Z. Priel and Y. Rabin, {\it J. Chem. Phys.}
{\bf 88}, 7111 (1988)


\refis{DE86} M. Doi and S.F. Edwards, {\it The Theory of Polymer
dynamics}, (Oxford University Press, Oxford, 1986)

\refis{DJ85} J. des Cloiseaux and  G. Jannink, {\it Les polym\`eres
en solution, leur mod\'elisation et leur structure},
(Editions de Physique, Paris, 1985)

\refis{DC73} J. des Cloizeaux, {\it Macromolecules}, {\bf 6}, 403 (1973)

\refis{DR86} R.M. Davis and W.B. Russel,
{\it J. Polym. Sci. B}, {\bf 24}, 511 (1986)

\refis{BJ92} J-L. Barrat, J-F. Joanny, P. Pincus,
{\it J. Physique II} {\bf 2}, 1531(1992)

\refis{DC95} A.V. Dobrynin, R.H. Colby and M. Rubinstein,
{\it Scaling theory of polyelectrolyte solutions"} (To appear
in {\it Journal de Physique II},  1995).

\refis{Napper} D.H. Napper,
     {\it  Polymeric stabilization of colloidal dispersions} 
   (Academic Press, New-York, 1983)

\refis{entanglements} Y.H. Lin, {\it Macromolecules},
{\bf 20}, 3080 (1987).

\refis{BF90} F.S. Bates, G.H. Fredrickson, 
{\it Ann. Rev. Phys. Chem.} {\bf 41}, 525 (1990)

\refis{GE77} P.G. de Gennes, 
  {\it  Rivista del Nuovo Cimento} {\bf 7}, 363 (1977)

\refis{FI82} M. Fixman, {\it J. Chem. Phys.} {\bf 76}, 6346 (1982).

\refis{FK} R.M. Fuoss, A. Katchalsky, S. Lifson. {\it Proc.
Natl. Acad. Sci. USA} {\bf 37}, 579 (1951)

\refis{GE79} P.G. de Gennes, {\it Scaling concepts in
polymer physics} (Cornell University Press, Ithaca, 1979)

\refis{Brochardetal} J. Hayter, G. Jannink, F. Brochard-Wyart and
P.G. de Gennes, {\it J. Phys. Lett.}, {\bf 41}, L-451 (1980)

\refis{Onsager} L. Onsager, {\it Ann. N.-Y. Acad. Sci.},
{\bf 51}, 627 (1949)

\refis{GP76} P.G. de Gennes, P. Pincus, R.M. Velasco, F. Brochard,
{\it J. Physique} {\bf 37}, 1461 (1976)

\refis{HT95} B.Y. Ha and D. Thirumalai, {\it Macromolecules}
{\bf 28}, 577 (1995).

\refis{HM86} J-P. Hansen, I.R. McDonald, {\it Theory of simple liquids}
(Academic Press, New-York, 1986)

\refis{Flory} P.J. Flory,  {\it Principles of Polymer Chemistry},
 (Cornell University Press, Ithaca, 1954)

\refis{LL} L.Landau, E.M. Lifchitz, {\it Physique Statistique},
(Mir, Moscow, 1967)

\refis{dgbrush}  P.G. de Gennes, {\it  Macromolecules} $\bf{13}$,
1069 (1980)

\refis{AL79} S. Alexander, {\it  J. Physique}  $\bf{38}$, 983 (1977)

\refis{Hill} T.L Hill,
{\it Introduction to Statistical Thermodynamics}, Dover (1986)

\refis{HK} W. Hess and R. Klein, {\it Adv. Phys.} {\bf 32 }, 173 (1983)

\refis{annealed} E. Raphael and J.F. Joanny, {\it Europhys. Lett.},
{\bf 13}, 623 (1990)

\refis{Ise} J. Yamanaka, H. Matsuoka, 
H. Kitano and N. Ise, {\it J. Coll. Int. Sci.}, {\bf 134}, 92 (1990).

\refis{Donnan} F.Donnan, Z.Physik.Chem.A,{\bf 168}, 369 (1934);
 F.Donnan, E.Guggenheim, Z.Physik.Chem.A, {\bf 122}, 346 (1932)

\refis{PI91} P. Pincus, Macromolecules, {\bf 24}, 2912 (1991)

\refis{BE} V. Borue, I. Erukhimovich , Macromolecules, {\bf 21}, 3240 (1988)

\refis{KK82} A.R. Khokhlov and  K.A. Khachaturian, {\it Polymer}
{\bf 23} (1982) 1793.

\refis{KH80} A.R. Khokhlov,  {\it J. Phys. A },  {\bf 13}, 979 (1980)

\refis{KK48} W. Kuhn, O. K\"unzle and A. Katchalsky
{\it Helv. Chim. Acta} {\bf 31}, 1994 (1948)

\refis{KP88} M.W. Kim and D.G. Peiffer, {\it Europhys. Lett.}, 
{\bf 5}, 321 (1988).

\refis{LE80} L. Leibler, {\it Macromolecules}, {\bf 13}, 1602 (1980).

\refis{MA68}G.S.  Manning, {\it J. Chem. Phys. } {\bf 51}, 954 (1969).

\refis{MA81} G.S. Manning, {\it J. Phys. Chem.}, {\bf 85}, 1506 (1981).

\refis{MR94} J. Ray and  G.S. Manning, {\it Langmuir }, {\bf 10}, 2450 (1994).

\refis{MU94} M. Muthukumar,  {\it Macromol. Theory Simul.}, {\bf 3}, 71 (1994)

\refis{MW83} G. Maret and  G. Weill, {\it Biopolymers}, {\bf 22},
 2727 (1983).

\refis{OD77}   T. Odijk, {\it J. Polym. Sci} {\bf 15}, 477 (1977)

\refis{OH78} T. Odijk and A.C. Houwaart, 
{\it J. Polym. Sci} {\bf 16}, 627 (1978)

\refis{OS71} F. Oosawa, {\it Polyelectrolytes} (Dekker, Ney-York, 1971).

\refis{PV77} P. Pfeuty, R. Velasco and P.G. de Gennes,
 {\it J. Physique. Lett} {\bf 38}, L5 (1977)

\refis{PH52} D.T. Pals and J.J. Hermans, 
{\it Rev. Trav. Chim.} {\bf 71}, 433 (1952)

\refis{OdijkMandel} M. Mandel and T.Odijk,
{\it Ann. Rev. Phys. Chem.} {\bf 35}, 75 (1984)

\refis{PF78} P. Pfeuty, {\it J. Physique}  {\bf 39} C2-149. (1978) 

\refis{RE68} P.M. R\'esibois, {\it Electrolyte Theory}, 
(Harper and Row, New-York, 1968)

\refis{RE94} W.F. Reed, {\it J. Chem. Phys.} {\bf 101}, 2515 (1994)

\refis{Schmitz} K.S. Schmitz in {\it Macro-ion Characterization}
K.S. Schmitz, Ed.  (American Chemical Society, Washington, 1994)

\refis{Reed} See for exemple the contribution by 
W.F. Reed in {\it Macro-ion Characterization}
K.S. Schmitz, Ed.  (American Chemical Society, Washington, 1994)

\refis{RS89}  W.B. Russel, D.A. Saville and  W.R. Schowalter.  
  {\it Colloidal dispersions}
  (Cambridge University Press, Cambridge, 1991).

\refis{RC94} M. Rubinstein, R.H. Colby and A.V. Dobrynin
{\it Phys. Rev. Lett. }  {\bf 73}, 2776 (1994)

\refis{Panyukov} S.V. Panyukov, {\it Sov. Phys. JETP} {\bf 71}, 372 (1990)

\refis{liqcryst} P.G. de Gennes and J. Prost,
      {\it The physics of liquid crystals} 
 (Oxford University Press, Oxford,  1993)

\refis{SF77}J. Skolnick and M. Fixman, {\it Macromolecules}
{\bf 10}  944. (1977).

\refis{SH82} J.D. Sherwood,
{\it J. Chem. Soc. Faraday Trans. 2}, {\bf 79} (1982).

\refis{JL91} J.-F. Joanny 
and L. Leibler, {\it J. Physique} {\bf 51}, 545 (1990)

\refis{SC91}  M. Schmidt, {\it Macromolecules}  {\bf 24}, 5361 (1991)

\refis{Hermans} J.J. Hermans, {\it J. Polym. Sci.} {\bf 18}, 527 (1955)

\refis{TA61} Tanford, {\it Physical Chemistry of Macromolecules},
(Wiley, New-York, 1961)

\refis{WO78} C. Wolff, {\it Journal de Physique - Colloques}
{\bf 39}, C2-169 (1978).

\refis{LiWitten} H. Li and T.A. Witten,
{\it Fluctuations and persistence length 
of charged flexible polymers}, (preprint 1995).

\refis{JP95} B. Jonsson, T. Peterson and B. Soderberg,
{\it J. Phys. Chem.}, {\bf 99}, 1251 (1994)

\refis{AS} M. Abramovitz and I.A. Stegun,
{\it Handbook of mathematical functions}, Dover 1965.

\refis{HiggsJoanny} P.G. Higgs and J.-F. Joanny,
{\it J. Chem. Phys.} {\bf 94 }, 1543 (1991)

\refis{frenkel} D. Frenkel, H. Lekkerkerker and A. Stroobants, {\it Nature},
 {\bf 332}, 822 (1988)

\refis{Turnerjoanny} M.S. Turner and J.-F. Joanny, 
{\it J. Phys. Chem.} {\bf 97}, 4825 (1993)

\refis{BJ94} J.-L. Barrat and J.-F. Joanny, {\it J. Phys. II} {\bf 4}, 
 1089 (1994)

\refis{Milner} S.T. Milner, {\it Science} {\bf 251}, 905 (1991)

\refis{tanakabenedek} T. Tanaka, L.O. Hocker and G.B. Benedek,
   {\it J.  Chem. Phys.}  {\bf 59}, 5151 (1973)

\refis{Yeomans} H.T. Dobbs and J. M. Yeomans,
{\it Mol. Phys.} {\bf 80}, 877 (1993)

\refis{Robbins} M.J. Stevens and M.O. Robbins,
{\it Europhys. Lett.}, {\bf 12}, 81 (1990)

\refis{Markorabin} J.F. Marko and Y. Rabin, 
    {\it Macromolecules}  {\bf 25} 1503 (1992)
     
\refis{OD94} T. Odijk, 
    {\it Macromolecules}  {\bf 27} 4998 (1994)

\refis{grosbergkhokhlov} A.Y. Grosberg and A.R. Khokhlov,
    {\it  Statistical physics of macromolecules}
(AIP Press, New York, 1994)

\refis{semenov} 
A.N. Semenov and A.R.  Kokhlov.
 {\it  Soviet Physics Uspekhi} {\bf 31}, 988 (1988)

\refis{Mandel} See e.g. the contribution  by M. Mandel 
in {\it  Encyclopedia of polymer science and engineering}
 Herman F. Mark et al editors,   (Wiley, New-York, 1990)

\refis{Marcelja} See e.g. the contribution  by S. Marcelja
in {\it Liquids at Interfaces}, J. Charvolin, J-F. Joanny and J. Zinn-Justin
editors (North-Holland, Amsterdam, 1990)

\refis{SafranChaikin} See e.g. the contribution by  P. Chaikin in
{\it  Physics of complex and supermolecular fluids}  S.A. Safran
   and N.A. Clark editors, (Wiley, New York, 1987)

\refis{Cedlak} M. Sedlak, {\it Macromolecules} {\bf 28}, 793 (1995)

\refis{KardarKantor} Y. Kantor, H. Li and  M. Kardar,
{\it Phys. Rev. Lett.}
{\bf 69}, 61 (1992)

\refis{tanakahanshibayama} M. Shibayama, T. Tanaka and C.C. Han,
{\it J. Phys. IV} {\bf 3}, 25 (1993)

\refis{dormidontova} E.E. Dormidontova, I.Y. Erukhimovich
and A.R. Khokhlov, {\it Colloid and Polymer Science} {\bf 272}, 1486 (1994)

\refis{OostwalOdijk} M. Oostwal and T. Odijk, {\it Macromolecules}
{\bf 26}, 6489 (1993)

\refis{IS85} J.N. Israelachvili,  {\it Intermolecular and Surface Forces} 
(Academic Press, London, 1985)

\refis{nyrkova2} I.A. Nyrkova, A.R. Khokhlov and M. Doi, 
{\it Macromolecules} {\bf 27}, 4220 (1994)

\refis{nyrkova3} A.R. Khokhlov and I.A. Nyrkova, {\it Macromolecules}
{\bf 25}, 1493 (1992)

\refis{KS94} M.J. Stevens and K. Kremer, {\it Phys. Rev. Lett.}
{\bf 71}, 2229 (1993); {\it Macromolecules}, {\bf 26}, 4717 (1993);
{\it J. Chem. Phys.}, in press (1995)

\refis{Kremerdunweg} B. D\"unweg, M.J. Stevens and K. Kremer in
{\it Computer Simulation in Polymer Physics}, K. Binder ed.
(Oxford University Press, Oxford, in press 1995)

\refis{Borisov}   E.B. Zhulina, O.V. Borisov and  T.M. Birshtein,
{\it Macromolecules} {\bf 28}, 1491 (1995)

\refis{ZB92}  E.B. Zhulina, O.V. Borisov and  T.M. Birshtein,
{\it J. Phys. II} {\bf 2}, 63 (1992)

\refis{WP87} T.A. Witten and P. Pincus, {\it Europhys. Lett.},
 {\bf 3}, 315 (1987)

\refis{OD79} T. Odijk, {\it Macromolecules}, {\bf 12}, 688 (1979)

\refis{pouyeteisenberg} H. Eisenberg and J. Pouyet,  {\it J. Pol. Sci.}
{\bf 13}, 85 (1954)

\refis{fuoss1} R.M. Fuoss, {\it Discuss. Farad. Soc.}, {\bf 11}, 125 (1951)

\refis{CS} J. Melenkevitz, K.S. Schweizer and J.G. Curro,
{\it Macromolecules} {\bf 26}, 6190 (1993)

\refis{Jannink} See e.g. the contribution by G.  Jannink in 
   {\it Physics and Chemistry of Aqueous Ionic Solutions}
 M. Bellisent-Funel and G.W.  Neilson editors,
  (Dordrecht,  Reidel, 1987)

\refis{woodward} C.E. Woodward and B. J\"onsson,
{\it Chemical Physics} {\bf 155}, 207 (1991)

\refis{Madden} H. L\"owen, J-P. Hansen and P.A. Madden,
{\it J. Chem. Phys.}  {\bf 98}, 3275 (1993)

\refis{JP80} J.F.Joanny and P.Pincus,
{\it Polymer} {\bf 21}, 274 (1980)

\refis{mobility} F.Wall and P.Grieger,
{\it J. Chem. Phys.} {\bf 20}, 1200 (1952)

\refis{Belloni} M.Olvera de la Cruz, L.Belloni, M.Delsanti, J.P.Dalbiez, O.Spalla and M.Drifford
{\it Precipitation of highly charged polyelectrolyte solutions in the presence of multivalent salts}
(submitted to Journal of Chemical Physics 1995)

\refis{Wittmer} J.Wittmer, A.Johner and J.F.Joanny,
{\it Precipitation of polyelectrolytes in the 
presence of multivalent salts} (Journal de Physique in press 1995)

\refis{Miklavic} S.Miklavic and S.Marcelja,
{\it J.Phys.Chem.} {\bf 92}, 6718 (1988)

\refis{Indiens} S.Misra,S.Varanasi and P.Varanasi,
{\it Macromolecules} {\bf 22}, 5173 (1989)

\refis{RubinsteinJoanny} R.Colby, M.Rubinstein, A.Dobrynin and J.F.Joanny,
{\it Elastic modulus and equilibrium swelling of polyelectrolyte gels} (preprint 1995)

\refis{Candau} R.Skouri, F.Schosseler, J.P.Munch and S.J.Candau
{\it Macromolecules} {\bf 28}, 197 (1995)

\refis{Kajijphys} K.Kaji, K.Urakawa, T.Kanata and R.Kitamaru,
{\it J.Physique (France)} {\bf 49}, 993 (1988)

\refis{nyrkova} I.Nyrkova, N.SHusharina and A.Khokhlov,
{\it Polymer preprints} {\bf 34}, 939 (1993)

\refis{Candau2} A.Moussaid, F.Schosseler, J.P.Munch and S.J.Candau,
{\it J.Physique II (France)} {\bf 3}, 573 (1993)

\refis{Schmidt} S.Foerster, M.Schmidt and M.Antonietti,
{\it Polymer} {\bf 31}, 781 (1990)

\refis{Hagerman} P.Hagerman,
{\it Biopolymers} {\bf 20},251 (1981)

\refis{leibleriliopoulos} M.Perreau, I.Iliopoulos and R.Audebert,
{\it Polymer} {\bf 30}, 2112 (1989)

\refis{khokhlov1} A.Khokhlov and E.Kramarenko,
{\it Macromolecular Theory and Simulations} {\bf 3}, 45 (1994)

\refis{adsorption} R.Varoqui,
{\it J.Physique II (France)} {\bf 3}, 1097 (1993)

\refis{Petitpas} X.Auvray, R.Anthore, C.Petipas, J.Huguet and M.Vert,
{\it J.Physique (France)} {\bf 47}, 893 (1986)

\endpage

\references
\endreferences

\endpage

\figurecaptions

\parindent=0pt

{\bf figure 1}: schematic representation of the blob structure 
of a charged polymer chain. The chain is gaussian up to the scale $\xi_e$, 
stretched on larger length scales.

{\bf figure 2}: schematic "state diagram" for a polyelectrolyte solution,
in the plane $c$ (polymer concentration), $n$ (concentration of
monovalent added salt). The boundaries between the different 
regions of the plane indicate crossovers, not sharp transitions.
Region I (dilute) is  discussed in sections 2, 3 and 6.1. Regions 
II,III and IV (semidilute)  are discussed in section 6.2 and 6.3. Region V
 (concentrated) is discussed in section 6.5.

{\bf figure 3}: Phase diagram of a dense polyelectrolyte solution in a bad
solvent.  Upper part: phase diagram in the concentration, temperature
plane for a fixed salt concentration. Lower part: phase diagram as
a function of salt concentration and temperature for a given polymer concentration.

{\bf figure 4}: schematic representation of the 
reduced viscosity as a function of polymer concentration $c$
at a fixed salt concentration $n$. 

\endfigurecaptions
\end